\documentclass[useAMS,usenatbib,usegraphicx]{mn2e}

\usepackage{pslatex,color,tabularx,float,multicol,caption}
\usepackage{amsmath,amsfonts,amssymb,amsxtra,eufrak,url}

\def\idm#1{{\mbox{\scriptsize #1}}}

\def\vec#1{{\pmb #1}}

\newcommand{\g}{\mbox{g}}
\newcommand{\yr}{\mbox{yr}}
\newcommand{\cm}{\mbox{cm}}
\newcommand{\s}{\mbox{s}}
\newcommand{\K}{\mbox{K}}
\newcommand{\au}{\mbox{AU}}
\newcommand{\Cs}{\mbox{C}_{\idm{s}}}

\newcommand{\msun}{\mbox{M}_{\odot}}
\newcommand{\Rsun}{\mbox{R}_{\odot}}
\newcommand{\GammaL}{\Gamma_{\idm{L}}}
\newcommand{\Gammac}{\Gamma_{\idm{c}}}
\newcommand{\Teff}{\mbox{T}_{\idm{eff}}}
\newcommand{\Tirr}{\mbox{T}_{\idm{irr}}}
\newcommand{\mE}{\mbox{M}_{\oplus}}

\newcommand{\OmegaK}{\Omega_{\idm{K}}}

\newcommand{\gz}{g_{\idm{z}}}
\newcommand{\zinfty}{z_{\infty}}
\newcommand{\Ps}{P_{\idm{s}}}
\newcommand{\Fs}{F_{\idm{s}}}
\newcommand{\Ts}{T_{\idm{s}}}
\newcommand{\Tb}{T_{\idm{b}}}
\newcommand{\kappas}{\kappa_{\idm{s}}}
\newcommand{\tauab}{\tau_{\idm{ab}}}
\newcommand{\kB}{k_{\idm{B}}}
\newcommand{\mH}{m_{\idm{H}}}

\title[Migration of two planets in a disc and formation of MMRs]
{On the migration of two planets in a disc and the formation of mean motion resonances}
\author[Cezary Migaszewski]{Cezary Migaszewski$^{1,2}$\thanks{E-mail:migaszewski@umk.pl}\\
$^{1}$Institute of Physics and CASA$^*$, Faculty of Mathematics and Physics, University of Szczecin, Wielkopolska 15, 70-451 Szczecin, Poland\\
$^{2}$Centre for Astronomy, Faculty of Physics, Astronomy and Informatics, Nicolaus Copernicus University, Grudziadzka 5, 87-100 Toru\'n, Poland}
\begin{document}
%
\date{Accepted 2015 July 28.  Received 2015 June 26; in original form 2015 May 7}
\pagerange{\pageref{firstpage}--\pageref{lastpage}} \pubyear{2012}
\maketitle
\label{firstpage}

\captionsetup[figure]{labelfont=bf,font=small}

\begin{abstract}

We study the dynamics of a system of two super-Earths embedded in a protoplanetary disc. We build a simple model of an irradiated viscous disc and use analytical prescriptions for the planet-disc interactions which lead to migration. We show that depending on the disc parameters, planets' masses and their positions in the disc, the migration of each planet can be inward or outward and the migration of a two-planet system can be convergent (which may lead to formation of a first order mean motion resonance, MMR) or divergent (a system moves away from MMR). We performed $3500$ simulations of the migration of two-planet systems with various masses and initial orbits. Almost all of them end up as resonant configurations, although the period ratios may be very distant from the nominal values of a given MMR.
We found that almost all the systems resulting from the migration are periodic configurations.

\end{abstract}


\section{Introduction}

It is widely accepted that planets form around young stars in protoplanetary discs at distances larger than the ones observed in evolved systems. Gravitational interactions between a planet and a disc excites density waves in the disc. The existence of the waves results in a non-radial component of the force acting on the planet and, as a consequence, the planet migrates inwards \citep{Goldreich1979,Goldreich1980}. Since these pioneering works, significant progress has been made in the theory of the planet-disc interactions \citep[e.g.,][]{Tanaka2002,Paardekooper2010,Paardekooper2011}, showing that the direction as well as the rate of the migration depend on the planets' masses and the disc properties in a complex way.

The smooth convergent migration of the planets together with the eccentricity damping naturally leads to systems with $P_2/P_1 \approx (p+1)/p$ \citep{Lee2002,Snellgrove2001,Papaloizou2005}. Therefore, one could expect that the observational sample of multi-planet systems is dominated by resonant configurations. Nevertheless, \cite{Fabrycky2014} showed that a histogram of period ratios of systems discovered by the {\em Kepler} mission does not reveal clear maxima at the positions of the MMRs. 

There are several explanations of those features presented in the literature. One of the proposed mechanisms are the stochastic forces acting on the planets. The forces might origin from the turbulence in a disc. \cite{Nelson2005} showed that the turbulent density fluctuations can be larger than the spiral wakes excited by the planets of a few Earth masses. The stochastic forces might also result from the interactions between the planets and planetesimals \citep[e.g.,][]{Chatterjee2014}. The interactions can disrupt a MMR if a planetesimal disc has a mass of $\gtrsim 0.2\times$ the mass of the planets.
Stochastic forces can be also incorporated in a model heuristically \citep[e.g.,][]{Rein2012}. In such an approach, one can parametrize the migration, circularization and stochastic force. For certain values of the parameters, it is possible to reconstruct the histogram of the period ratios. Nevertheless, \cite{Hands2014} showed that increasing magnitude of the turbulence does not have to reduce significantly the number of resonant systems with respect to all the systems, but it rather reduces the survival rate of the systems.

Dissipation due to the planet-star tidal interaction is another mechanism which could be responsible for moving the systems out of resonances \citep{Papaloizou2010,Papaloizou2011,Batygin2013,Delisle2014a,Delisle2014b}. The tidal dissipation of the mechanical energy causes the inward migration of a planet, which is (in a two-planet system) faster for the inner planet.  Nevertheless, the tidal dissipation is fast enough only for short-period planets \citep[$P \lesssim 10\,$days, this value depends on the type of planets,][]{Lee2013}.
The divergent migration can also result from the interaction between a planet and a wake produced by another planet in the same system \citep{Podlewska-Gaca2012,Baruteau2013}. This mechanism is efficient if at least one of the planets opens a partial gap. 

\cite{Goldreich2014} show that the capture of planets into a MMR is permanent if the equilibrium eccentricity $e_{\idm{eq}} < (m/M)^{1/3}$, where $m$ and $M$ are masses of the planet and the star, respectively. The value of $e_{\idm{eq}}$ results from the balance between excitation of $e$ due to passing through MMR and damping the eccentricity due to planet-disc interaction. If the damping is not strong enough, $e_{\idm{eq}}$ can be too high and the capture into MMR is only temporary. In such a case resonant configurations should be rare. On the other hand, as shown by \cite{Deck2015} although a given system resides in a MMR only some limited time, shortly after leaving one MMR the system goes into another MMR. 

In this work we present another scenario of shifting systems away from MMRs. We show that in a standard 1+1D $\alpha-$disc model with a realistic opacity law, the disc can be divided into several regions of convergent and divergent migration. During the evolution of the disc, these regions move inwards at the viscous time-scale $\tau_{\idm{vis}}$. The planets migrate in the disc at rates which are in general different from $\tau_{\idm{vis}}$. Therefore, during the migration the period ratio can decrease (the system may be trapped into MMR) or increase (the system moves out of MMR) depending on the current positions of the planets in the disc. The final state of the system (in particular, the period ratio) depends on the masses and initial orbits of the planets. Variety of final period ratios can result from the migration. One of the possibilities are systems with $P_2/P_1 \approx (p+1)/p$ for $p=1, 2, 3$ or~$4$. Systems with the period ratios between the nominal values of two neighbouring first-order MMRs (like 2:1 and 3:2) are also possible. The migration can also lead to formation of hierarchical systems of period ratios $\gtrsim 10$. The common feature of almost all these systems is that they evolve periodically and at least one of the two resonant angles oscillates. Moreover, almost all systems with $P_2/P_1 \lesssim 2.12$ have two resonant angles which librate.

In section~2 we introduce a disc model. In section~3 we present analytical prescriptions of the planet-disc interactions and study the migration rate as a function of planet's mass and the size of its orbit. In section~4 we discuss the migration of a single planet. In the next section, the evolution of a two-planet system is studied. We show that sizes and positions of the regions of convergent and divergent migration depend on the planets' masses and evolutionary state of the disc. In section~6 we discuss main features of the sample of $3500$ systems. Section~7 is devoted to limitations of the model. Conclusions are given in Section~8.

\section{A model of a disc}

We consider a geometrically thin, viscous, axially symmetric disc. We assume that the planets do not affect the disc evolution and that the vertical and radial structures of the disc are independent \citep[1+1D methodology,][]{Garaud2007}. Below we describe the model.

\subsection{The diffusion equation}

In order to study the migration of planets down to very small distances from the star, we use the angular velocity profile different from usually used Keplerian profile $\OmegaK = (G \, M / r^3)^{1/2}$, where $G$ is the gravitational constant and $r$ is a distance from the star. In the so called boundary layer \citep{Pringle1981}, $\Omega$ decreases rapidly from a value close to $\OmegaK$ down to the rotational angular velocity of the star $\Omega_* \ll \OmegaK$. The boundary layer width $L$ is much smaller than the star's radius $R_*$. Assuming $\Omega_* = 0$, we chose
\begin{equation}
\Omega(r) = \OmegaK \biggl[ 1 - \left(\frac{R_*}{r}\right)^{\frac{1}{2}} \biggr].
\label{eq:Omega}
\end{equation}
This particular form of the angular velocity is convenient when one wants to apply it to the diffusion equation, which then reads: 
\begin{equation}
\frac{\partial \Sigma}{\partial t} = 3 \left( 1 - \zeta \right) \frac{\partial^2}{\partial r^2} \left( \bar{\nu} \, \Sigma \right) + \frac{9}{2 \, r} \left( 1 - \frac{1}{3} \zeta \right) \frac{\partial}{\partial r} \left( \bar{\nu} \, \Sigma \right),
\label{eq:diffusion2}
\end{equation}
where $\zeta \equiv (4/3) (R_*/r)^{1/2}$ and $\bar{\nu}$ is the turbulent viscosity coefficient averaged over the vertical axis $z$ and $t$ is time. Far from the star $\zeta \to 0$ and Eq.~(\ref{eq:diffusion2}) tends to usually used formula. 

\subsubsection{Photoevaporation}

The right-hand side of Eq.~(\ref{eq:diffusion2}) needs to be completed with a photoevaporation term $\dot{\Sigma}_w$. We follow \cite{Matsuyama2003}, but change slightly their $\dot{\Sigma}_w$ in order to have smooth transition between region of the disc from which the material is removed to the region of gravitationally bounded gas. We have
\begin{equation}
\dot{\Sigma}_w = 
\left\{
\begin{array}{l l}
\dot{\Sigma}_0 \, \left( \frac{r}{r_g} \right)^{-5/2}, & r \geq r_g, \\
\dot{\Sigma}_0 \, \exp\bigg[-\left( \frac{r - r_g}{\beta \, r_g} \right)^2\bigg], & (1 - \beta) r_g \leq r < r_g.
\end{array}
\right.
\label{eq:photoevaporation}
\end{equation}
where $\dot{\Sigma}_0 = 1.257 \times 10^{-13} T_{\idm{II},2}^2 \, \Phi_{41}^{1/2} m_0^{-3/2} \, \g\,\cm^{-2}\,\s^{-1}$, $T_{\idm{II},4}$ is the temperature of ionized HII atmosphere above the disc divided by $10^4\,\K$, $\Phi_{41}$ is the ionizing photons luminosity divided by $10^{41} \s^{-1}$, 
$m_0 = M_*/\msun$ and $r_g = 7.26 \, m_0 \, T_{\idm{II},4}^{-1} \, \au$, . We adopted $\Phi_{41} = 1$, $\beta = 0.2$ and $T_{\idm{II},4} = 1$.
Equation~(\ref{eq:diffusion2}) together with $\dot{\Sigma}_w$ is being solved with the following boundary condition.
At $r = r_{\idm{inner}}$ we assume constant accretion rate ($\Sigma \, \bar{\nu} = \mbox{const}$), while the outer boundary condition reads $\Sigma(r_{\idm{outer}}) = 0$.

\begin{figure*}
\centerline{
\vbox{
\hbox{
\includegraphics[width=0.49\textwidth]{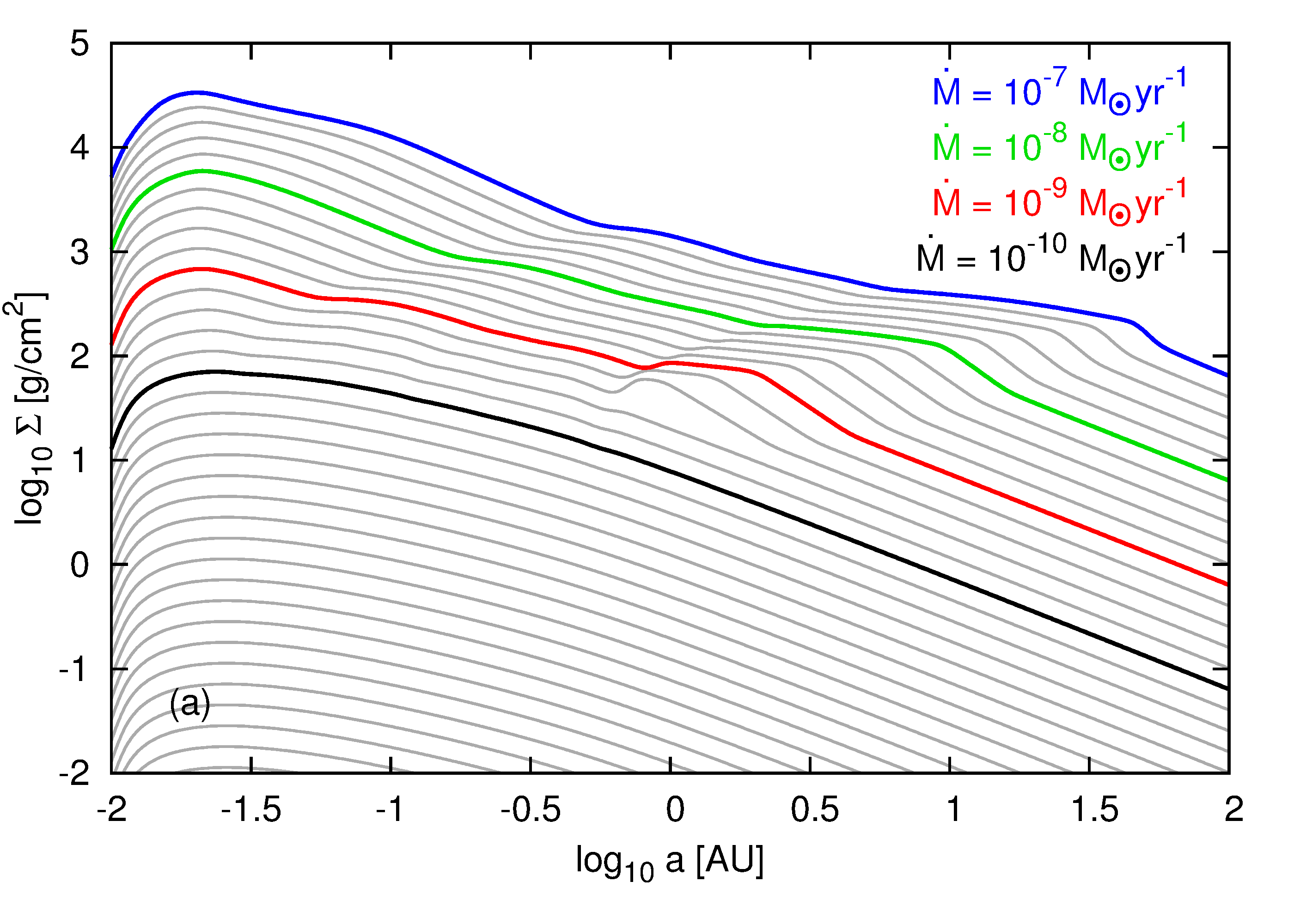}
\includegraphics[width=0.49\textwidth]{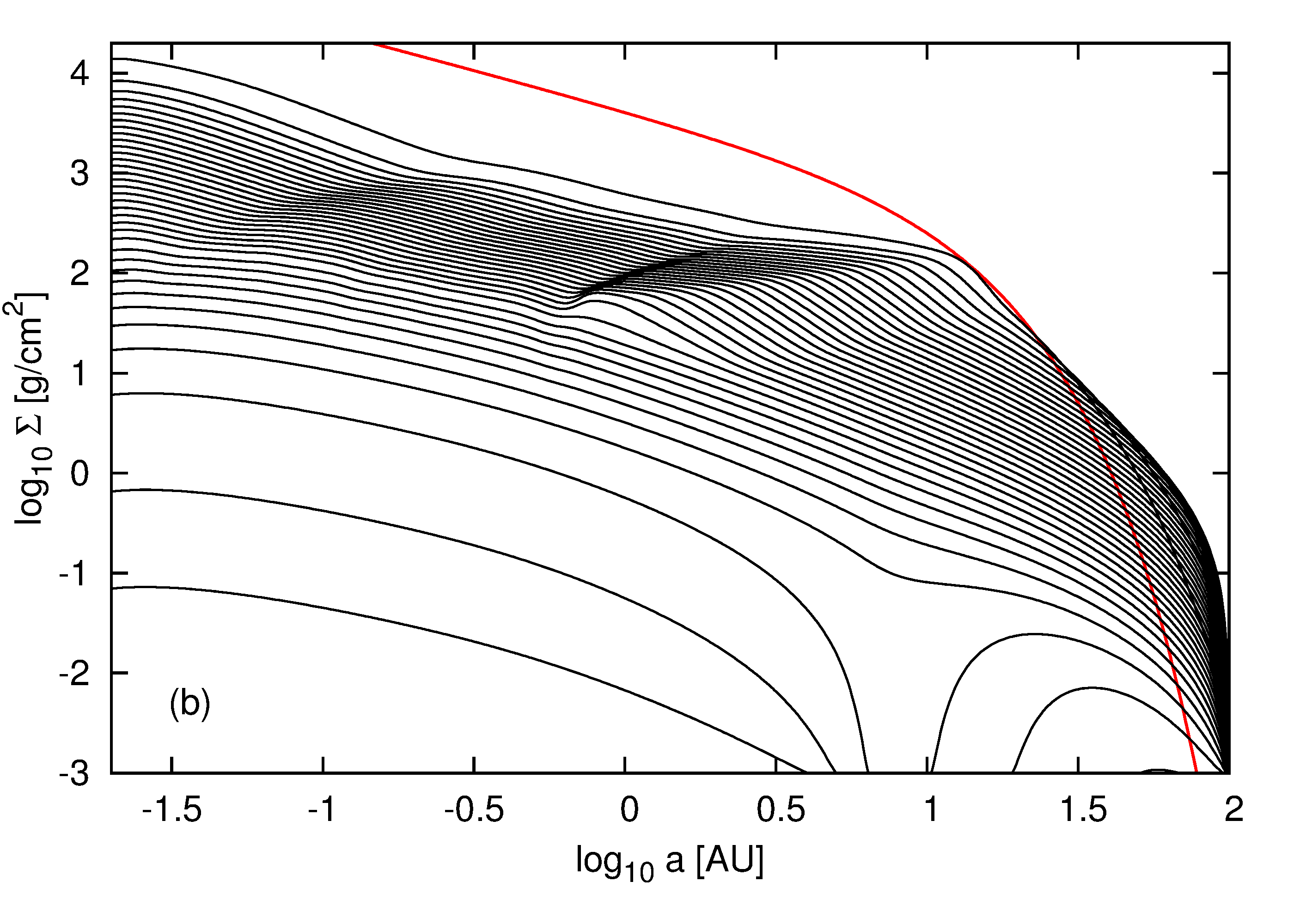}
}
}
}
\caption{{\em Panel (a)} Surface density profile for several values of constant $\dot{M}$ (labelled accordingly). {\em Panel (b)} Surface density evolution. Red curve is for the initial disc, black curves show $\Sigma-$profiles every $10^5$~years.}
\label{fig:Sigma_Temp_h_Mdot_const}
\end{figure*}

\subsection{The vertical structure equations}

To solve Eq.~\ref{eq:diffusion2}, one needs to know $\bar{\nu}$ at a given distance $r$ and a given time $t$. This is achieved by solving standard equations of the vertical structure. The only modification we make is that $\Omega(r)$ is given by Eq.~(\ref{eq:Omega}) instead of $\OmegaK$. The equations read as follows:
\begin{eqnarray}
\frac{d P}{d z} &=& -\rho \, \gz = -\rho \, \OmegaK^2 \, z \left( 1 + \frac{z^2}{r^2} \right)^{-\frac{3}{2}},\label{eq:dPdz}\\
\frac{d F}{d z} &=& D(r) = \rho \, \nu \left( r \, \frac{d \Omega}{d r} \right)^2 = \frac{9}{4} \, \rho \, \nu \, \OmegaK^2 \left( 1 - \zeta \right)^2,\label{eq:dFdz}\\
\frac{d T}{d z} &=& \nabla \frac{T}{P} \, \frac{d P}{d z}\label{eq:dTdz},
\end{eqnarray}
where $P, \rho, F$ and $T$ denote pressure, density, energy flux and temperature, respectively.
The viscosity coefficient is given by the $\alpha$ prescription \citep{Shakura1973,DAlessio1998}
\begin{equation}
\nu = \nu(z) = \frac{\alpha \, P(z)}{\rho(z) \, \Omega}, \quad \alpha = \mbox{const}.
\end{equation}
We assume that the energy is transported by radiation, convection and turbulence \citep{DAlessio1998}. The gradient
$
\nabla \equiv d  \, \ln T/d \, \ln P
$
is computed with the help of the mixing length theory \citep[we follow the approach of][]{Heinzeller2009}.

\subsubsection{Boundary conditions}

Following \citep{Papaloizou1999} we write the boundary conditions, $\Ps \equiv P(\zinfty)$, $\Fs \equiv F(\zinfty)$, $\Ts \equiv T(\zinfty)$, $F_0 \equiv F(z=0)$:
\begin{eqnarray}
\Ps &=& \frac{\OmegaK^2 \, H \, \tauab}{\kappas},\label{eq:Ps}\\
\Fs &=& \frac{3 \, \dot{M}}{8 \, \pi} \, \OmegaK^2 \left( 1 - \zeta \right)^2,\label{eq:Fs}\\
0 &=& 2 \, \sigma \left( \Ts^4 - \Tb^4 \right) - \frac{9 \, \alpha \, \OmegaK}{8 \, \kappas} \, \frac{\kB}{\tilde{\mu} \, \mH} \Ts \left( 1 - \zeta \right)^2 - \Fs,\label{eq:Ts}\\
F_0 &=& 0\label{eq:F0},
\end{eqnarray}
where $\zinfty$ is a height above which no energy is produced, the disc height $H$ is defined such that $\rho(H) = \rho(0) \, \exp(-1)$, $\dot{M}$ is the accretion rate, $\tauab$ is the optical depth above the disc, $\kappas$ is the opacity at $z=\zinfty$, $\Tb$ is the background temperature, $\tilde{\mu}$ is the mean molecular weight, $\kB$ means the Boltzmann constant and $\mH$ is the hydrogen atom mass.

\subsubsection{Stellar irradiation}

We use a simple approach to incorporate stellar irradiation of the disc, i.e., we modify temperature at the boundary $\Ts$ by adding $\Tirr$ in a forth power, i.e. $\Ts^4 := \Ts^4 + \Tirr^4$ \citep{Hueso2005}, where $\Tirr$ has a form of \citep{Ruden1991}:
\begin{equation}
\Tirr = T_* \bigg[ \frac{2}{3\,\pi} \left(\frac{R_*}{r}\right)^3 + \frac{1}{2} \left( \frac{R_*}{r} \right)^2 \left( \frac{H}{r} \right) \left( \frac{d \, \ln H}{d \, \ln r} - 1 \right)\bigg]^{\frac{1}{4}}.
\end{equation}
It is a common approach to assume $d\ln H / d\ln r = 9/7$ \citep[e.g.,][]{Mordasini2015}. This assumption seems to be a reasonable choice, because it is an equilibrium solution for a disc in regions dominated by the irradiation \citep{Chiang1997} and, on the other hand, $\Tirr \ll \Ts$ in regions dominated by the viscous heating. Nevertheless, the approximation does not have to be good in regions where the viscous heating and the irradiation are of similar magnitude.
We resign from this simplification and write $\Tirr$ as:
\begin{equation}
\Tirr =  T_* \bigg[ \frac{2}{3\,\pi} \left(\frac{R_*}{r}\right)^3 + \frac{1}{2} s_h \left( \frac{R_*}{r} \right)^2 \bigg]^{\frac{1}{4}},
\label{eq:Tirr}
\end{equation}
where $h \equiv H/r$ is the aspect ratio and $s_h \equiv r \, dh/dr$ is an unknown to be found in a self-consistent way.

\subsubsection{Opacity}

We use opacity tables from \cite{Semenov2003} instead of usually used formulae \citep{Ruden1991}. \cite{Semenov2003} computed $\kappa$ for variety of dust models (several models of the grain structure and topology as well as the chemical composition). We checked that the differences does not affect the disc profile significantly. In further study we use a model of composite spherical grains of typical mass fraction of metallic iron. 

\subsection{Solving the equations of the vertical structure}

We solve Eqs.~(\ref{eq:dPdz})-(\ref{eq:dTdz}) for a given $r$ and $\dot{M}$ starting from $z = \zinfty$ down to $z=0$. Our unknowns are $\zinfty$ and $s_h$.
A method to solve the equations is the following. We choose some value of $s_h$, let's call it $s_h^{\idm{(adopted)}}$. Once we have $s_h$ at a given radius $r$ for a given $\dot{M}$, we know $\Ts$. Then we find such $\zinfty$ for which $F_0 = 0$. Then all the physical quantities ($P, F, T, \rho$) are known as functions of $z$, which gives us also $H$. Repeating the process for radii $r - \delta r$ and $r + \delta r$ we find $s_h = s_h^{\idm{(computed)}}$. A proper solution of the vertical structure equations means $s_h^{\idm{(computed)}} = s_h^{\idm{(adopted)}}$.
In principle, there are three possible types situations for a given $\dot{M}$ and $r$, i.e., an irradiated disc, a shadowed disc and a non-unique solution. Figure~\ref{fig:Sigma_Temp_h_Mdot_const}a shows $\Sigma$-profiles for several values of $\dot{M} \leq 10^{-7}\,\msun\,\yr^{-1}$. For a reference, the disc of $\dot{M} = 10^{-9}\,\msun\,\yr^{-1}$ (red curve) is shadowed for $\log_{10}r[\au] \in [\sim 0, \sim 0.3]$.

\begin{figure*}
\centerline{
\vbox{
\hbox{
\includegraphics[width=0.49\textwidth]{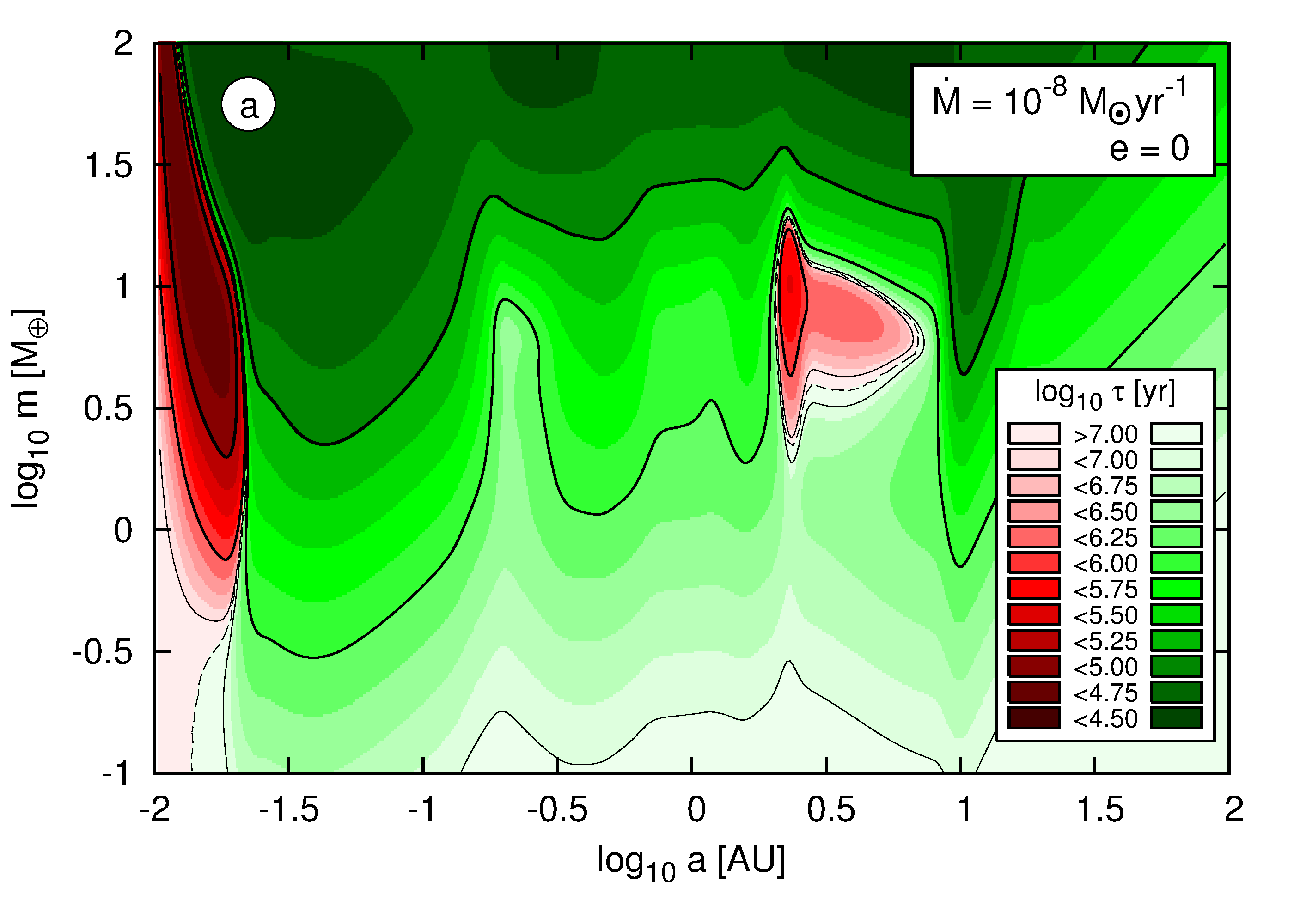}
\includegraphics[width=0.49\textwidth]{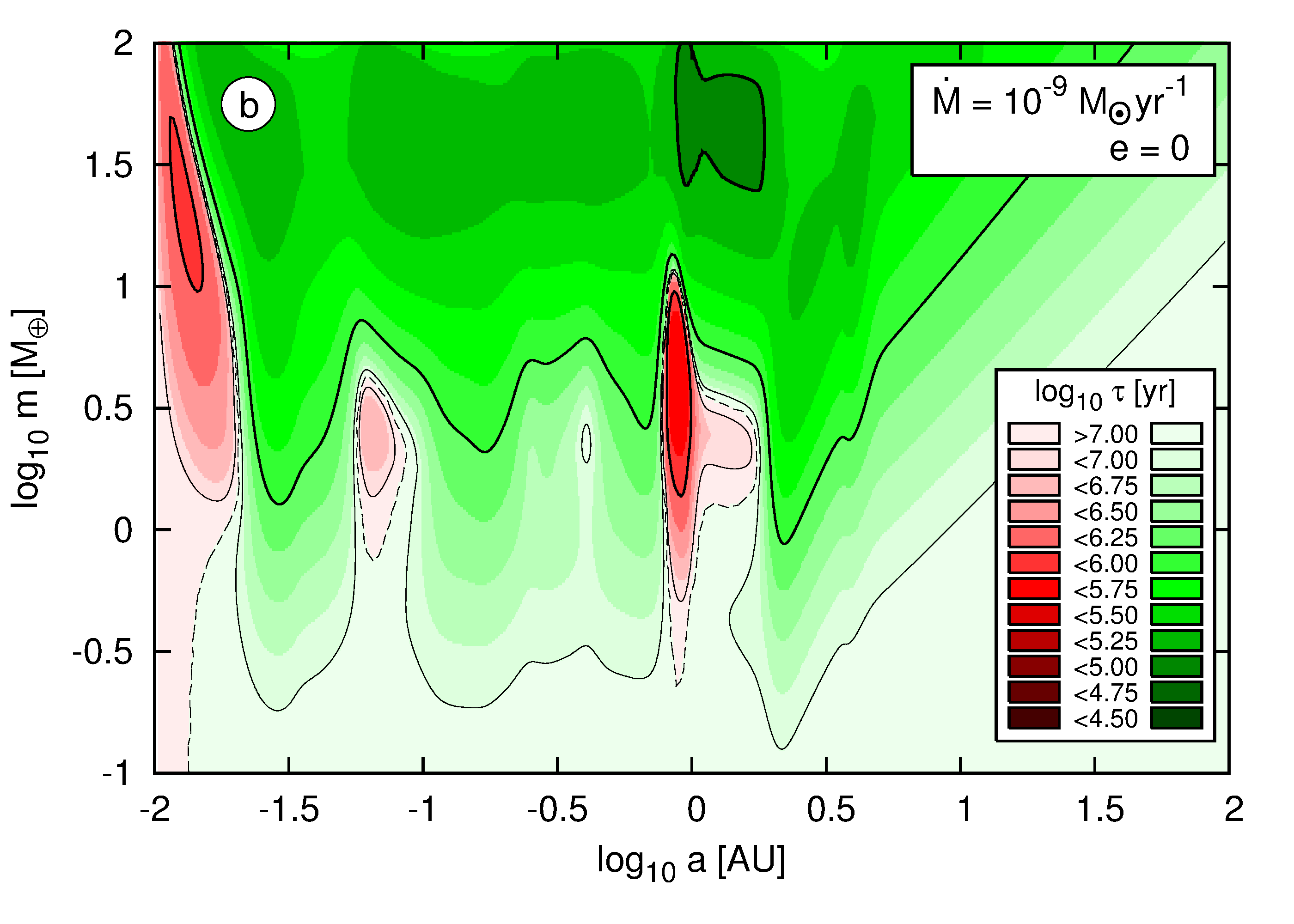}
}
\hbox{
\includegraphics[width=0.49\textwidth]{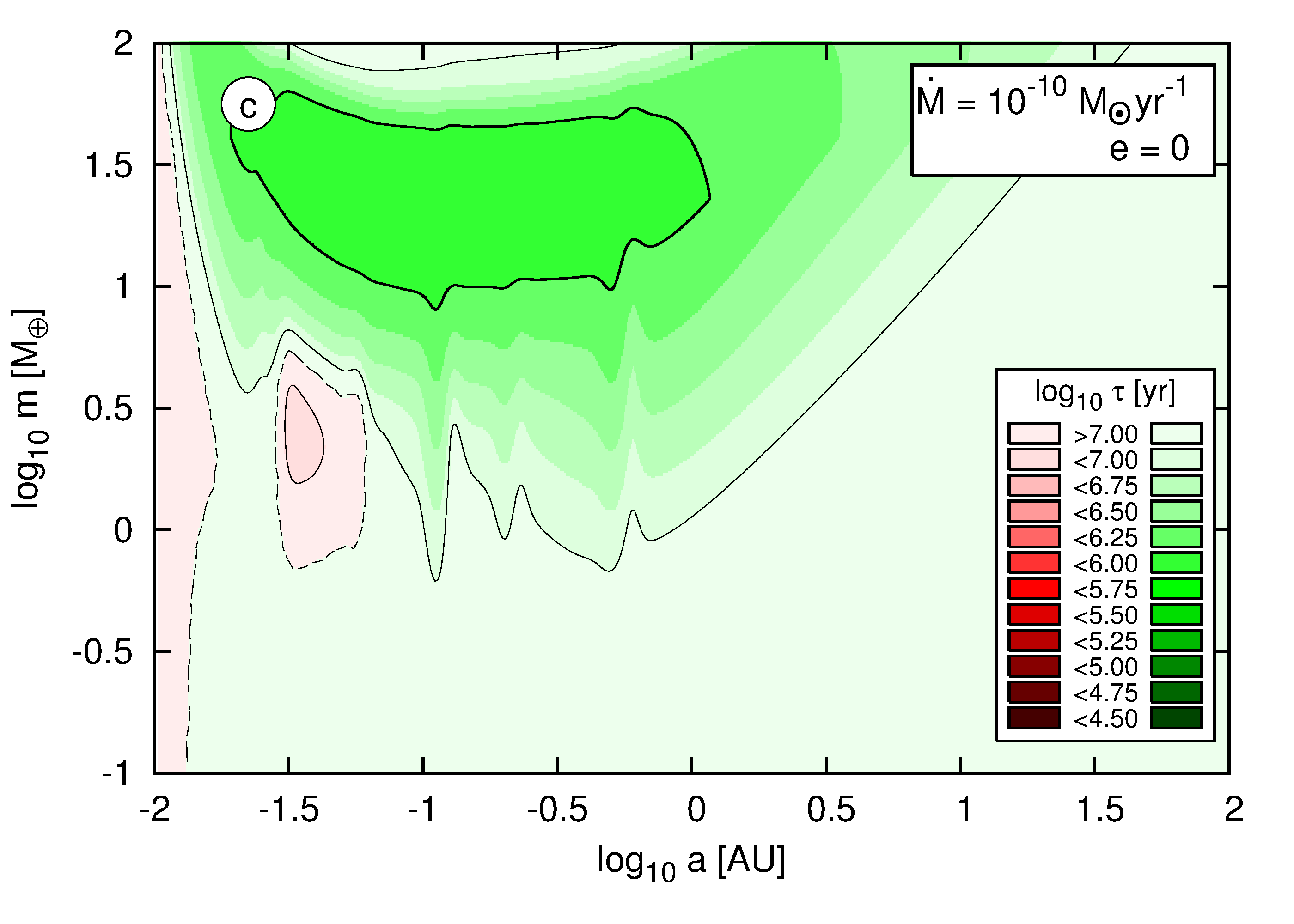}
\includegraphics[width=0.49\textwidth]{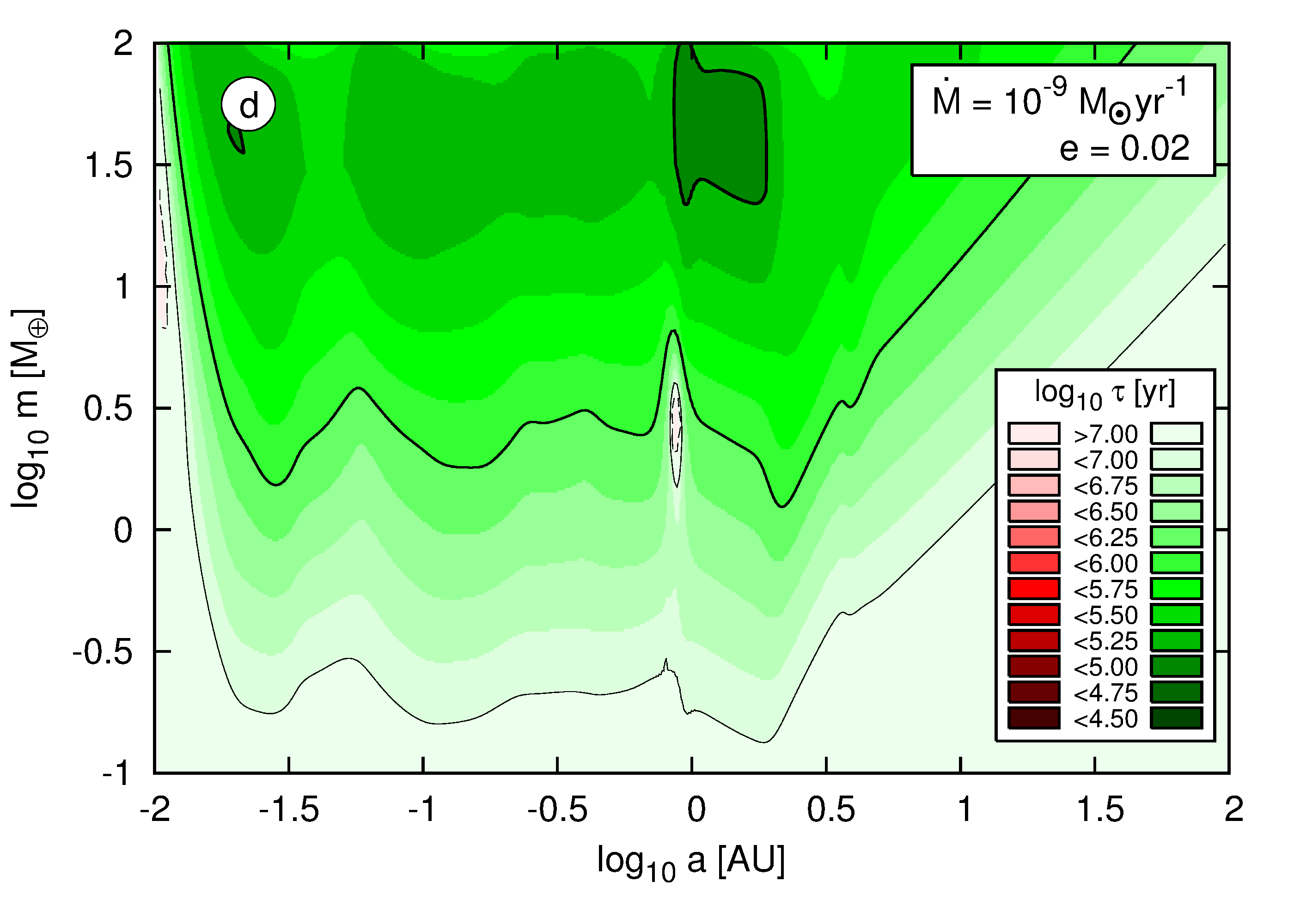}
}
}
}
\caption{The migration rate as a functions of $m$ and $r$. Green colour denotes the inward migration, while red is for the outward migration. The darker the colour is, the faster is the migration. The accretion rate and the eccentricity are given at each panel.
}
\label{fig:torque1}
\end{figure*}

\subsection{Time evolution of the disc}

To solve Eq.~(\ref{eq:diffusion2}) together with the photoevaporation term, Eq.~(\ref{eq:photoevaporation}), we need to know $\bar{\nu}$ at given $t$ and $r$. Nevertheless, it is not necessary to solve Eqs.~(\ref{eq:dPdz})-(\ref{eq:dTdz}) at each time-step of the evolution of the disc. For a given $r$, $\bar{\nu}$ is a function of $\dot{M}$ and because $\Sigma$ is a monotonically increasing function of $\dot{M}$ at a given radius, one can write $\bar{\nu} = \bar{\nu}(\Sigma, r)$. This way we close Eq.~(\ref{eq:diffusion2}) in a sense, that our only unknown is $\Sigma = \Sigma(t, r)$ and $\bar{\nu}$ is given by $\Sigma$ at given $r$ and $t$. One needs to solve the vertical structure equations once for different $r$ and $\dot{M}$ values taken from a grid ($\log_{10} r[\au] \in [-2,2]$ with a step of $0.005$, $\log_{10} \dot{M}[\msun \yr^{-1}] \in [-16,-6]$ with a step of $0.1$). Exact values of $\bar{\nu}$ for given $\Sigma$ are obtained with the help of the interpolation.

The disc model is computed for the following values of the parameters: $\alpha = 0.004$, $M_* = 1\,\msun$, $R_* = 2\,\Rsun$, the effective temperature of the star $\Teff = 4000 \K$.
The initial profile $\Sigma(t=0, r) \propto (r/a_0)^{-b} \, \exp[-(r/a_C)^{(2-b)}]$ is taken after \cite{Fortier2013},
where we chose $b = 0.8, a_0 = 5.0\,\au, a_C = 10.0\,\au$. The initial mass of the disc is $0.04\,\msun$, the inner radius $0.02\,\au$, and the outer radius $100\,\au$. The outer radius as well as values of $b, a_0$ and $a_C$ are not crucial for the evolution of the inner parts of the disc, $r \lesssim 10\,\au$. We integrate Eq.~(\ref{eq:diffusion2}) until the disc is dispersed. Figure~\ref{fig:Sigma_Temp_h_Mdot_const}b shows the evolution of $\Sigma$. Red curve is for the initial profile. Black curves show the profiles every $10^5\,\yr$. The shadowed disc region exists in first $\sim 2.3~$Myr of the evolution. The photoevaporation is the most effective for radii of a few $\au$. Similar evolution of $\Sigma$-profile (apart from the shadowed region) can be found in \citep{Mordasini2012}.

\section{Planet--disc interaction}

Because all the studies leading to analytical prescriptions of the torque are devoted to single planet--disc interaction (the density waves in a disc excited by one planet, affect only this planet, not the other planets embedded in the same disc) and finding the formulae for a general case is beyond the scope of this paper, we use here this approximation.
We also assume that the planet does not affect the disc evolution. It means that once we solved Eq.~\ref{eq:diffusion2}, we use the solution $\Sigma(t, r)$, $T(t, r)$, etc. to compute the force acting on a planet. The disc structure is given on a uniform grid of $\log_{10}r$ and with a constant output time-step of $5~$kyr. Values of $\Sigma$ and $T$ on the grid give local power indices $\gamma_1$ and $\gamma_2$ ($\Sigma \propto r^{-\gamma_1}$, $T \propto r^{-\gamma_2}$). The physical quantities at a given $r$ and $t$ are obtained with the help of the spline interpolation in $\log_{10} r$ and the linear interpolation in $t$.
 
The total torque acting on a planet moving in a circular orbit is a sum of the Lindblad torque $\GammaL$ and the corotation torque $\Gammac$, $\Gamma = \GammaL + \Gammac$. We use formulae from \citep{Paardekooper2011}, where they study non-isothermal Type~I migration. $\Gamma$ is a function of masses of the star and the planet as well as the disc parameters \citep[see][for details]{Paardekooper2011}. 

Once we compute $\Gamma$ for a planet of a given mass $m$ in an orbit of a given size $r$, the acceleration acting on the planet is given by
\begin{equation}
\vec{f}_T = -\frac{\dot{\vec{r}}}{2\,\tau_a}, \quad \tau_a = -\frac{\tilde{m} \, \sqrt{\mu \, a}}{2\,\Gamma} = -\frac{L_0}{2\,\Gamma},
\end{equation}
where $\tau_a$ is the time-scale of migration, $a$ is the semi-major axis, $\mu = G \, (M_* + m)$, $\tilde{m} = (1/M_* + 1/m)^{-1}$ is the reduced mass of the planet, $L_0$ is the angular momentum of the planet in circular orbit and $\dot{\vec{r}}$ is the astrocentric velocity of the planet.

Because in general the orbit is non-circular, we take into account the eccentricity waves \citep{Tanaka2004}.
The formulae in that paper are given in a specific reference frame. The planet starts from the pericenter and the size of the orbit is fixed. The force acting on the planet is given as an explicit function of time. The formulae cannot be used directly in a more general case. Assuming coplanar orbits, we rewrite their Eqs.~[38]--[40] as 
$F_r = A \, \left( 0.057 \, x + 0.176 \, y \right)$ and 
$F_{\phi} = A \, \left( -0.868 \, x + 0.325 \, y \right)$,
where $A = m \, M_*^{-2} \, \Cs^{-4} \, a^6 \, e \, \Sigma \, \Omega^6$, $e$ is the eccentricity and $\Cs$ is the isothermal sound speed.
Cartesian coordinates of the planet $(x,y)$ are given in the orbital reference frame (the $x$-axis points the pericenter).
The acceleration acting on the planet reads:
\begin{equation}
\vec{f}_e = F_{\phi} \, \frac{r}{L} \, \dot{\vec{r}} + \left( F_r - F_{\phi} \, \frac{\vec{r} \cdot \dot{\vec{r}}}{L} \right) \frac{\vec{r}}{r}, \quad L \equiv \| \vec{r} \times \dot{\vec{r}} \|.
\end{equation}
%

When the orbit is eccentric, the corotation torque is weaker than for a circular case \citep{Fendyke2014}. The corotation torque for the orbit of eccentricity $e$ is given by the torque for a circular orbit $\Gammac(0)$ multiplied by a factor of a value smaller than unity, $\Gammac(e) = \exp(-e/e_f) \, \Gammac(0)$, where $e_f = h/2 + 0.01$.

For masses of the planets $\gtrsim 10\,\mE$ and for not very massive disc, $\Sigma(1\au) \lesssim 100\,\g\cm^{-2}$, formulae of Type~I migration may not describe the disc-planet interaction correctly. Therefore, we use a simple analytic prescriptions from \citep{Dittkrist2014} for the transition between Type~I and Type~II migration formulae. 

The last term in our model is the gravitational force originating from an unperturbed disc. Papers devoted to the planet-disc interactions are usually focused on $a$ and $e$ evolution, thus the axially symmetric component of the potential from the disc is not taken into account as it leads only to the rotation of pericenter. When there are more planets in the disc, changes of the frequencies of the apsidal lines rotation affects the dynamics of the system. When $\Sigma(r)$-profile is given at time $t$, the central force acting on a planet has a form of $\vec{f}_r = -d(r) \, \vec{r}$,
where $d(r)$ stems from integration of the planet-disc interaction over a whole axially symmetric disc.

Formulae for $F_r$, $F_{\phi}$ and $\Gammac(e)$ depend explicitly on $e$ and $a$. The quantities have a meaning of the osculating Keplerian elements. Nevertheless, when the disc is massive and the planet is moving in a relatively wide orbit, the axially-symmetric component of the force from the disc can be a significant fraction of the central force from the star.
A planet in a circular orbit, i.e., $r(t) = \mbox{const}$, is moving with a super-Keplerian velocity. Without taking into account the additional force from the disc, one obtains $e > 0$ (also $a$ will be incorrect) when uses the standard formulae for the osculating Keplerian elements. The value of $e$ one obtains can be even $\gtrsim 0.1$, which leads to incorrect values of $\vec{f}_e$ and $\Gammac$.
Therefore, when computing $a$ and $e$ we take $\mu^* = \mu + d(r) \, r^3$ instead of $\mu$.

Finally, the equations of motion for $i$-th planet read as follows:
$\ddot{\vec{r}}_i = -\mu_i \, r_i^{-3} \, \vec{r}_i + \vec{f}_{\idm{p-p}}^{(i)} + \vec{f}_T + \vec{f}_e + \vec{f}_r$,
where 
\begin{equation}
\vec{f}_{\idm{p-p}}^{(i)} = -\sum_{j \neq i} \frac{G \, m_j}{\| \vec{r}_i - \vec{r}_j \|^3} \left( \vec{r}_i - \vec{r}_j \right) - \sum_{j \neq i} \frac{G \, m_j}{r_j^3} \, \vec{r}_j
\end{equation}
is the $N$-body component. The formulae given in this section are being used to integrate the equations of motion of the system of two planets. Before we go to this part of the analysis, in the next section we study how the migration rate depends on the planet's mass and position in the disc of given constant accretion rate.

\subsection{The migration rate as a function of $a$ and $m$}

Figure~\ref{fig:torque1}a,b,c present the migration rate for a planet in a circular orbit embedded in a disc of constant accretion rate of $10^{-8}, 10^{-9}$ and $10^{-10}\,\msun\,\yr^{-1}$, respectively. The migration rate is higher for larger $\dot{M}$ and lower for smaller $\dot{M}$. For $\dot{M}=10^{-8}$ and $10^{-9}\,\msun\,\yr^{-1}$, there exist regions of the outward migration in the shadowed part of the disc (they are limited to mass ranges between $\sim 1$ and $\sim 10\,\mE$) as well as close to the star (this regions extends for a whole range of planet's masses). For $\dot{M}=10^{-9}\,\msun\,\yr^{-1}$ the outward migration occurs also for $r \lesssim 0.1\,\au$. For lower accretion rate $\dot{M} = 10^{-10}\,\msun\,\yr^{-1}$ almost all the $(a, m)$-plane corresponds to the inward migration.
The migration rate depends also on $e$. Figure~\ref{fig:torque1}d presents $\tau_a$-map for $\dot{M}=10^{-9}\,\msun\,\yr^{-1}$ and $e=0.02$. All regions of the outward migration disappears (compare with Fig.~\ref{fig:torque1}b).

The existence of the regions of outward migration is important for planets growth during the migration \citep[e.g.,][]{Bitsch2014}. For a single-planet system, the eccentricity is damped to zero and this region exists. The result is that the planet is halted at the distance of a few~$\au$ until the mass grows enough to omit the trap. However, when two planets migrating in a disc form a mean motion resonance, their eccentricities are excited and the planets can omit the barrier even if their masses are small. This might lead to qualitative difference in the evolution of orbits as well as the mass growth of a single-planet system with respect to a multi-planet system. We leave this problem to future works.

\begin{figure}
\centerline{
\includegraphics[width=0.49\textwidth]{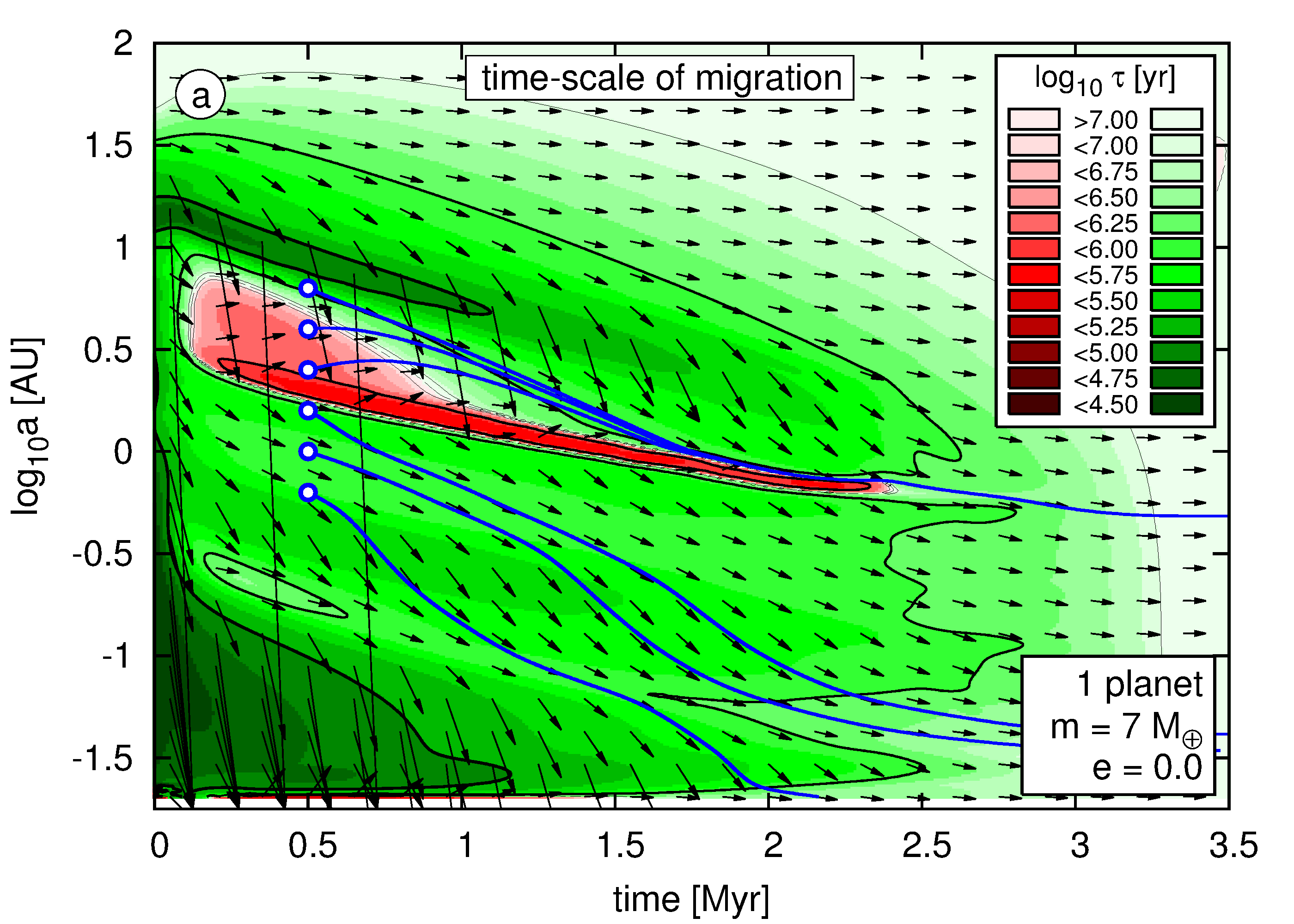}
}
\caption{{\em Colour-coded:} Migration rate of a single planet ($m = 7\,\mE$, $e=0$) embedded in the disc as a function of time and the distance from the star. {\em Vectors:} Velocity field of a planet. Blue curves show the evolution of the systems of chosen initial semi-major axes (white symbols at $t = 0.5~$Myr).}
\label{fig:one_planet_ecc_0}
\end{figure}

\begin{figure*}
\centerline{
\vbox{
\hbox{
\includegraphics[width=0.49\textwidth]{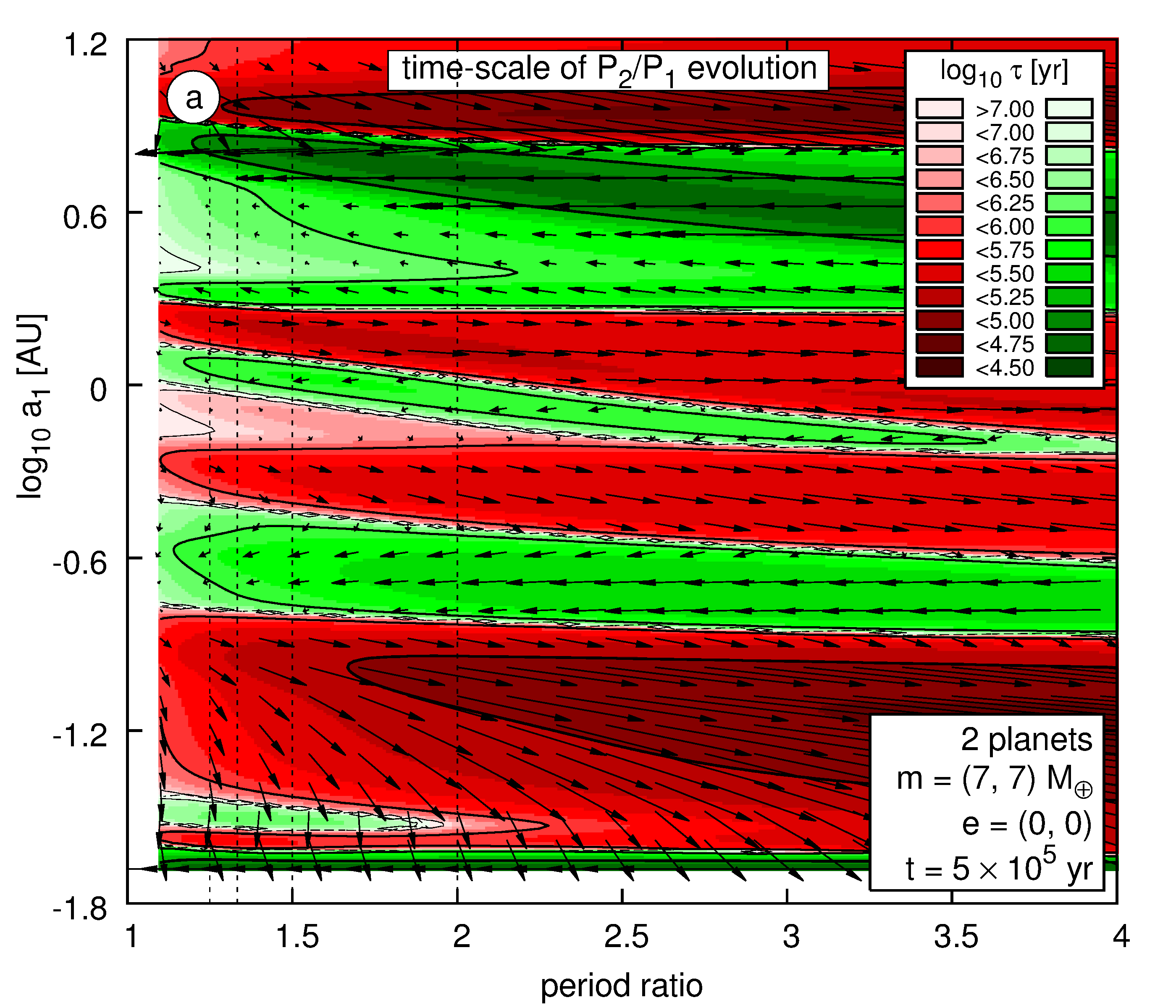}
\includegraphics[width=0.49\textwidth]{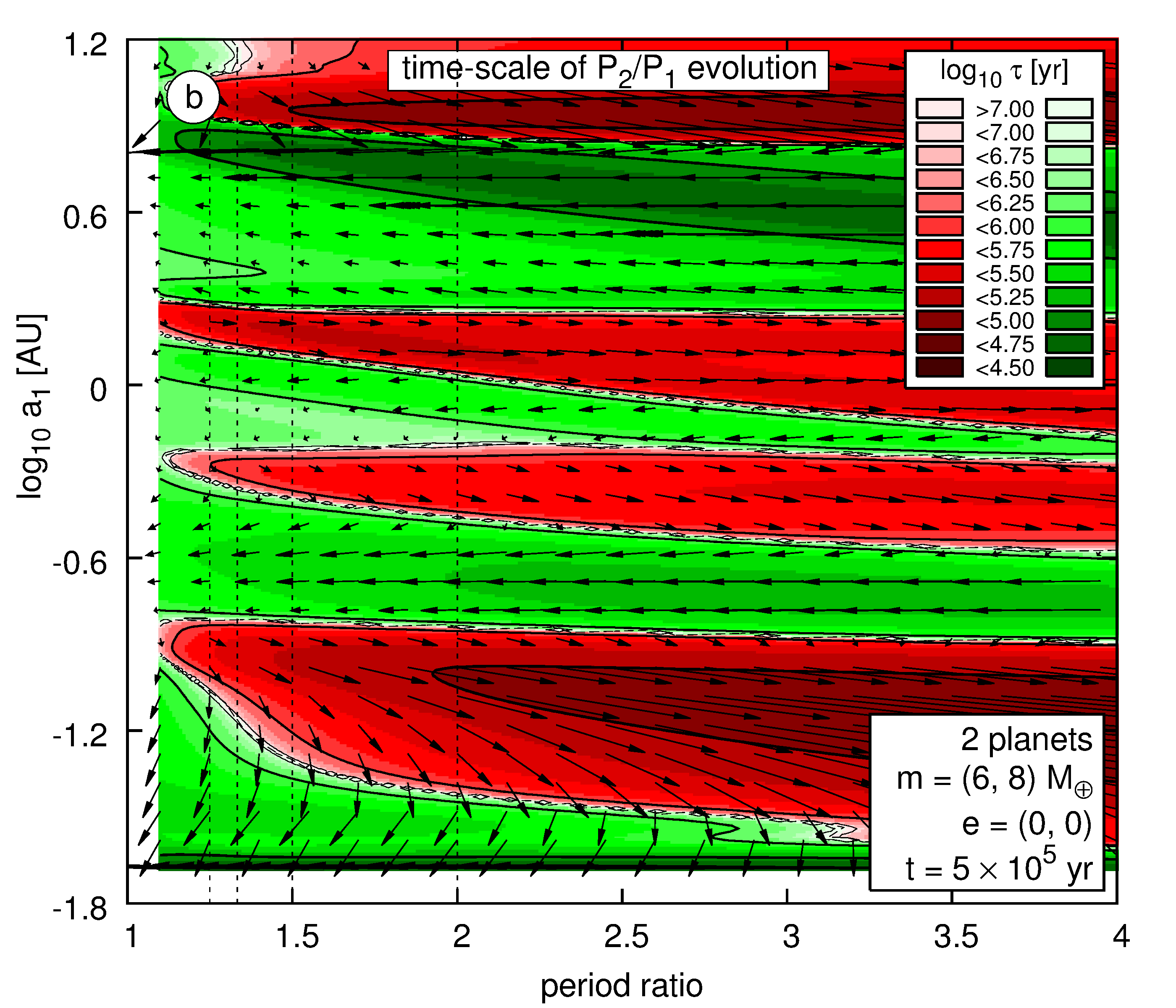}
}
\hbox{
\includegraphics[width=0.49\textwidth]{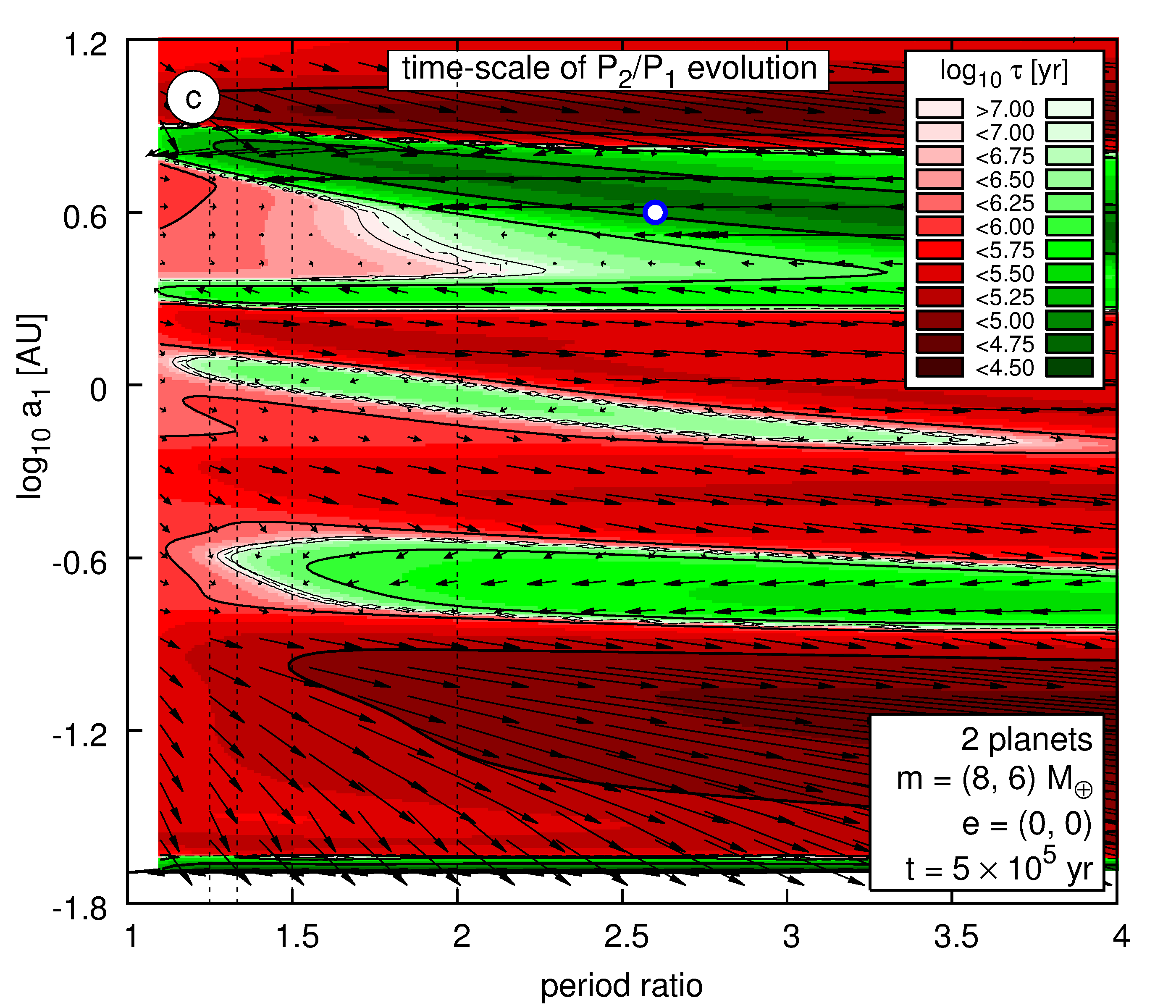}
\includegraphics[width=0.49\textwidth]{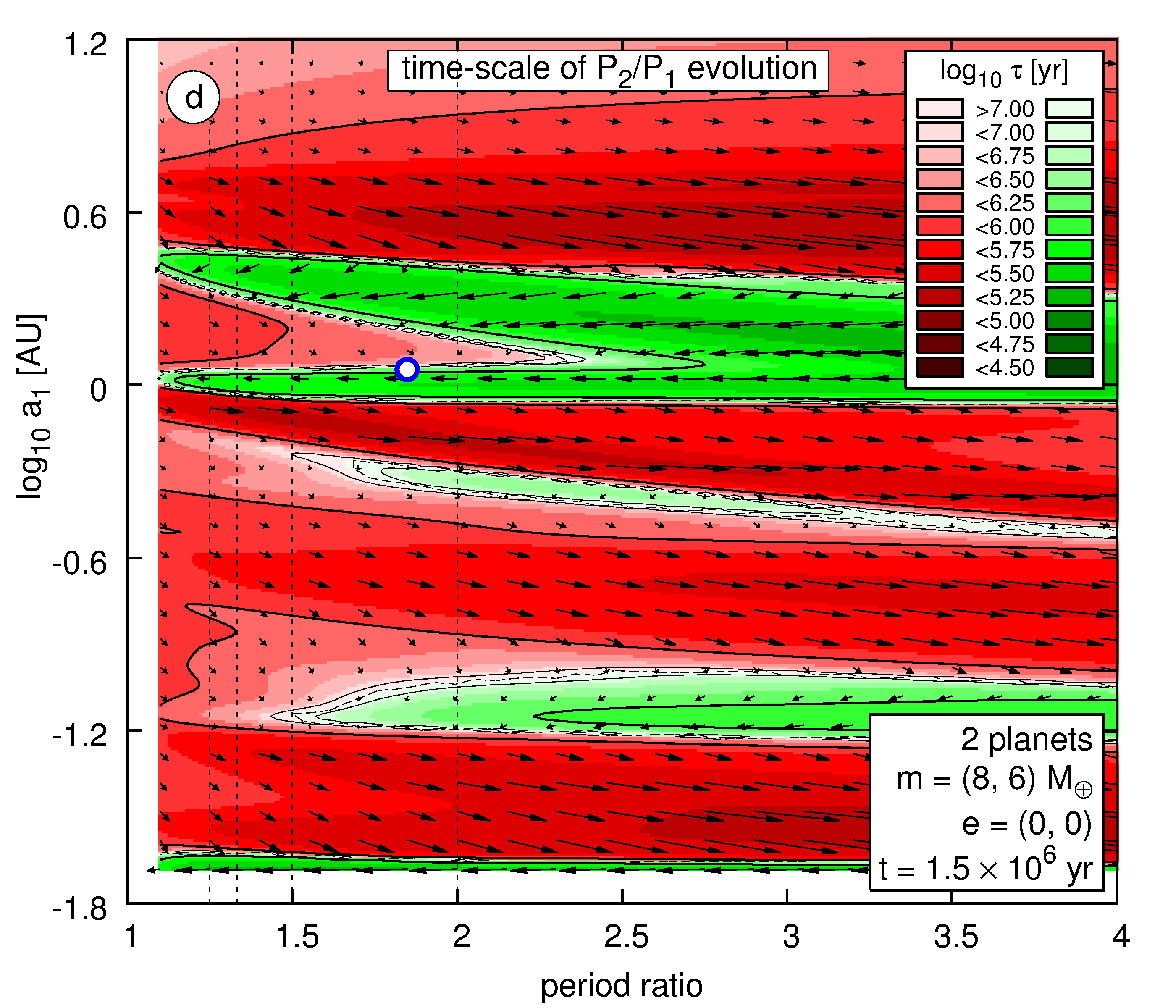}
}
}
}
\caption{{\em Colours:} Time-scale of the period ratio variation, $\tau_X$. {\em Arrows:} Vector field, i.e., the $x$-component of a vector is given by $\dot{X}$ multiplied by a time interval of $t_{\idm{vec}} = 1.5 \times 10^4\,\yr$, the $y$-component is given by $d(\log_{10}a_1)/dt$ multiplied by $t_{\idm{vec}}$. Green colour means that $\tau_X > 0$ (convergent migration), red colour denotes $\tau_X < 0$ (divergent migration). The darker the colour is, the larger is $|\tau_X|$. Masses of the planets are $m_1 = m_2 = 7\,\mE$ (panel a), $m_1 = 6\,\mE, m_2 = 8\,\mE$ (b) and $m_1 = 8\,\mE, m_2 = 6\,\mE$ (c). For all these plots  $t = 5\times10^5\,\yr$. Panel (d) shows the view of $\tau_X$-map for $m_1 = 8\,\mE, m_2 = 6\,\mE$ and $t = 1.5\times10^6\,\yr$.}
\label{fig:two_planet_7_7_0_0}
\end{figure*}

\section{Single planet migration}

In the previous section we discussed the migration time-scale of a planet of given mass in a disc of a given accretion rate. A series of migration maps obtained for decreasing $\dot{M}$ can be related to different evolution stages of the disc. The disc evolves approximately as a $\dot{M} = \mbox{const}$-disc, apart from the late stage of evolution, when the photoevaporation plays an important role. 
Figure~\ref{fig:one_planet_ecc_0} presents $\tau_a-$maps at $(t, \log_{10}a_1)-$plane for $m=7\,\mE$ and $e=0$. The outward migration region occupies relatively large range of radii and exists up to $\sim 2.3~$Myr of the disc life-time. 
The blue curves show the evolution of single-planet systems for different initial orbits.
A planet which starts the evolution in or above the red region of the map cannot reach orbits of sizes significantly smaller than $1\,\au$. The planet reaches the trap ($\Gamma = 0, \tau_a \to \infty$) after a short time and migrates further inward at the time-scale of the disc evolution, which is $\sim 1.6~$Myr. On the other hand, when a planet starts below the red area, it can migrate inward faster than the disc evolves. A planet starting at $r \sim 0.8\,\au$ can reach $r \sim 0.02\,\au$.

\begin{figure*}
\centerline{
\vbox{
\hbox{
\includegraphics[width=0.49\textwidth]{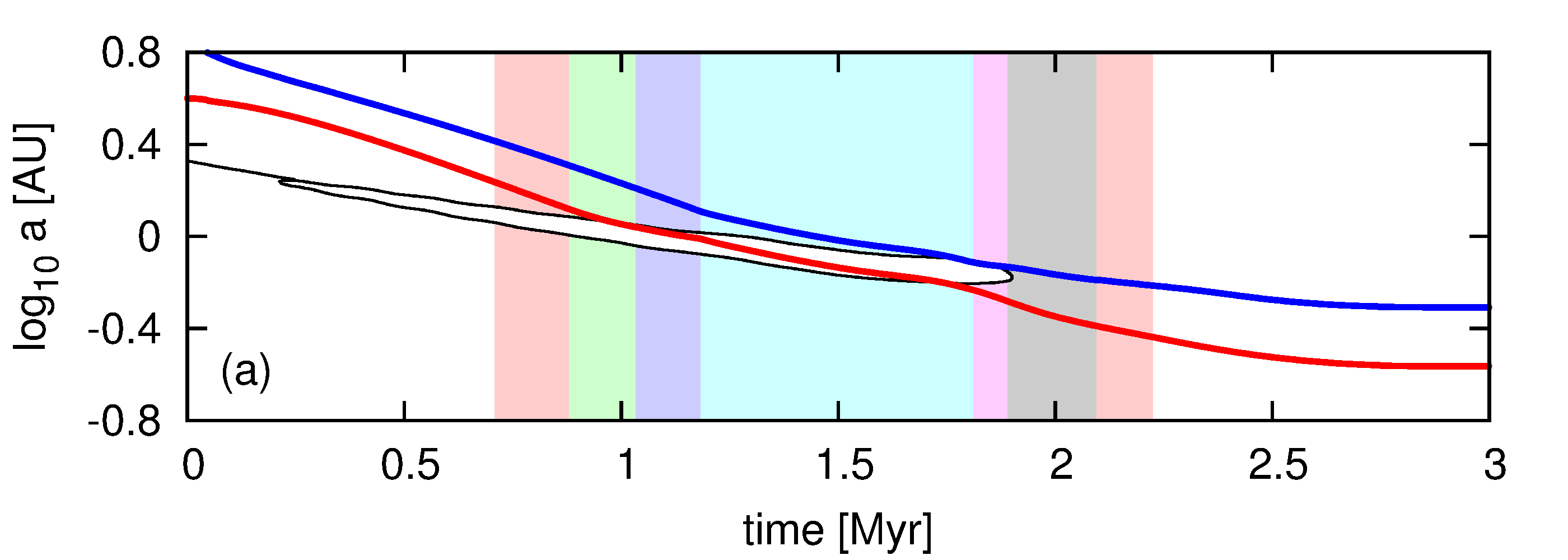}
\includegraphics[width=0.49\textwidth]{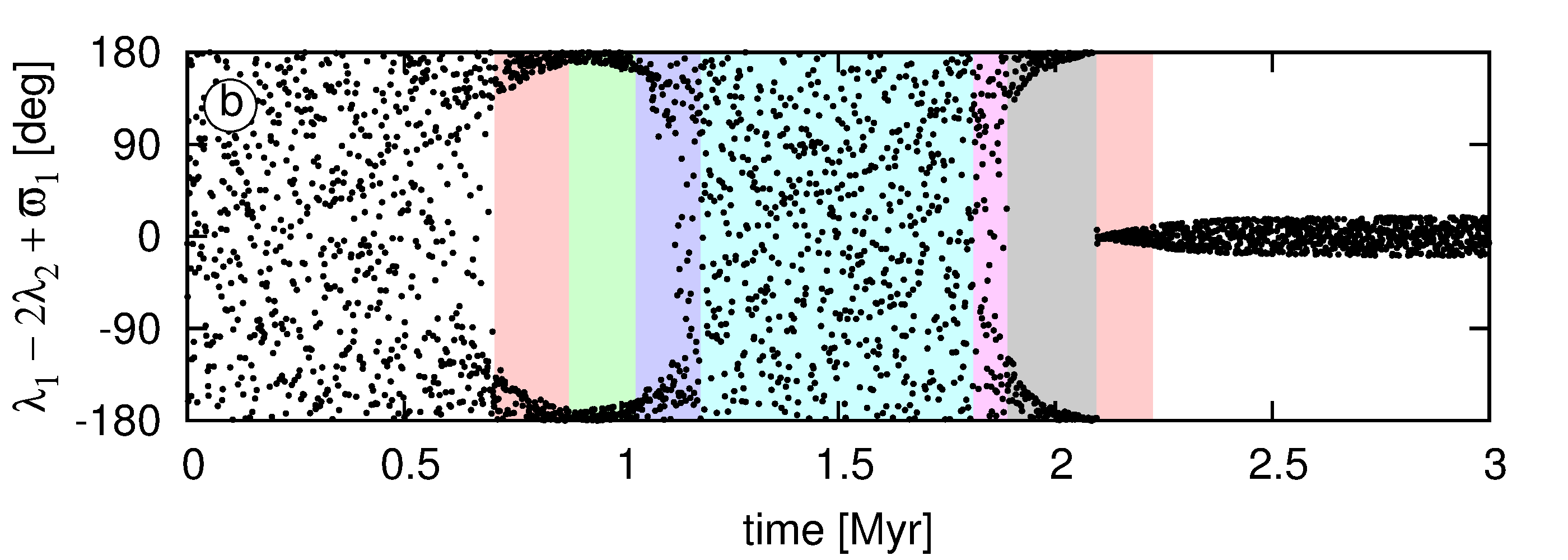}
}
\hbox{
\includegraphics[width=0.49\textwidth]{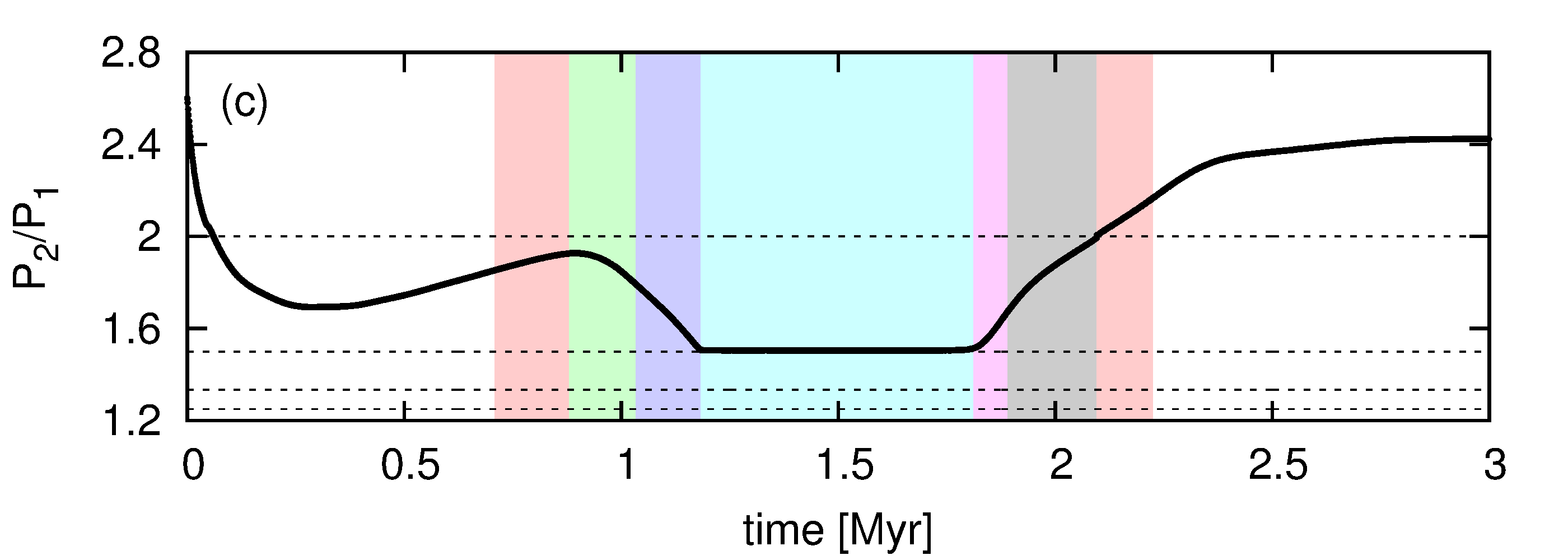}
\includegraphics[width=0.49\textwidth]{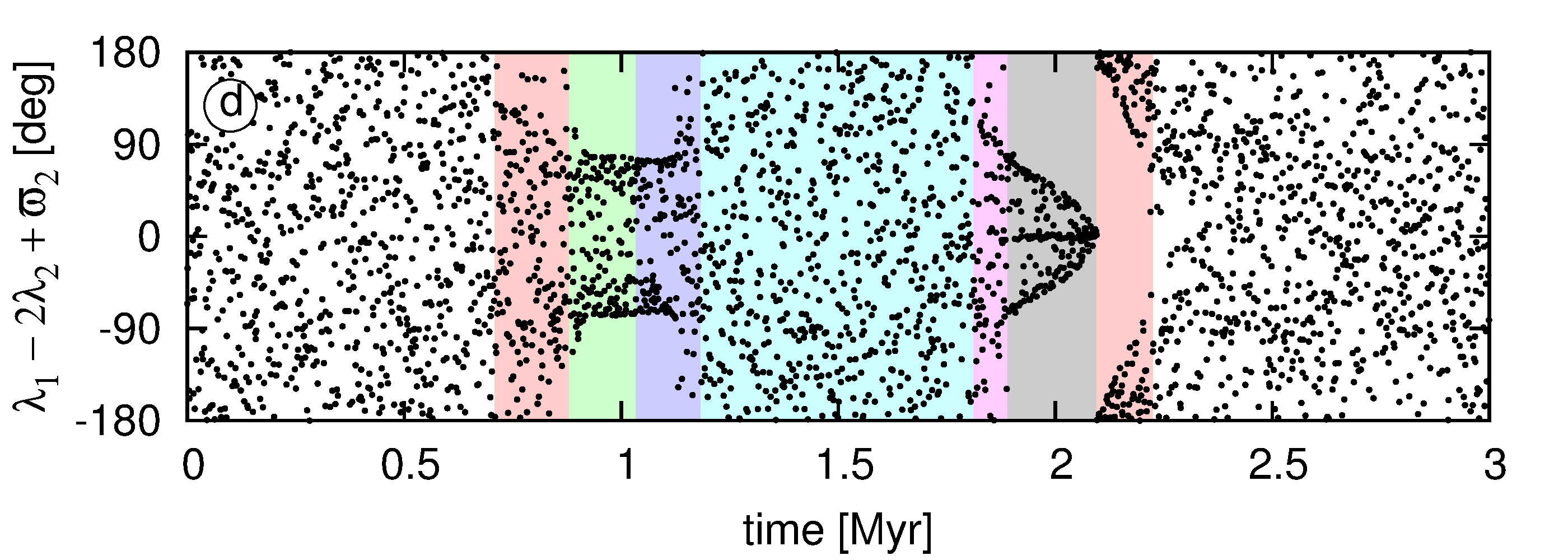}
}
\hbox{
\includegraphics[width=0.49\textwidth]{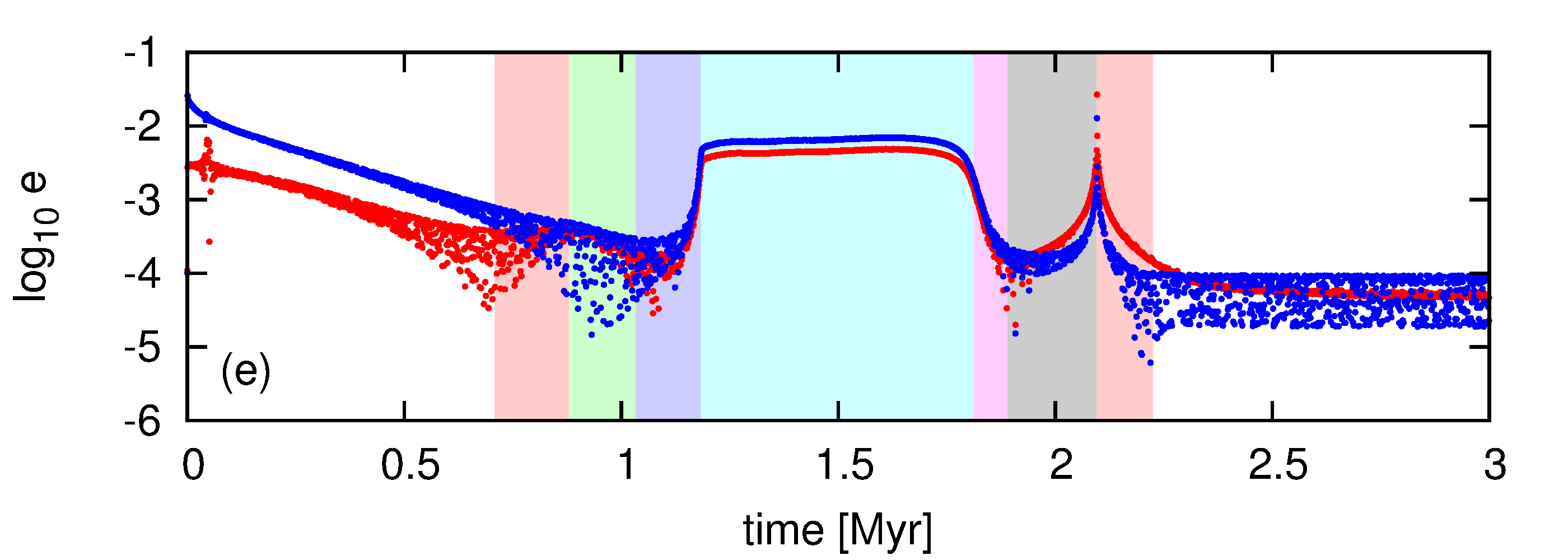}
\includegraphics[width=0.49\textwidth]{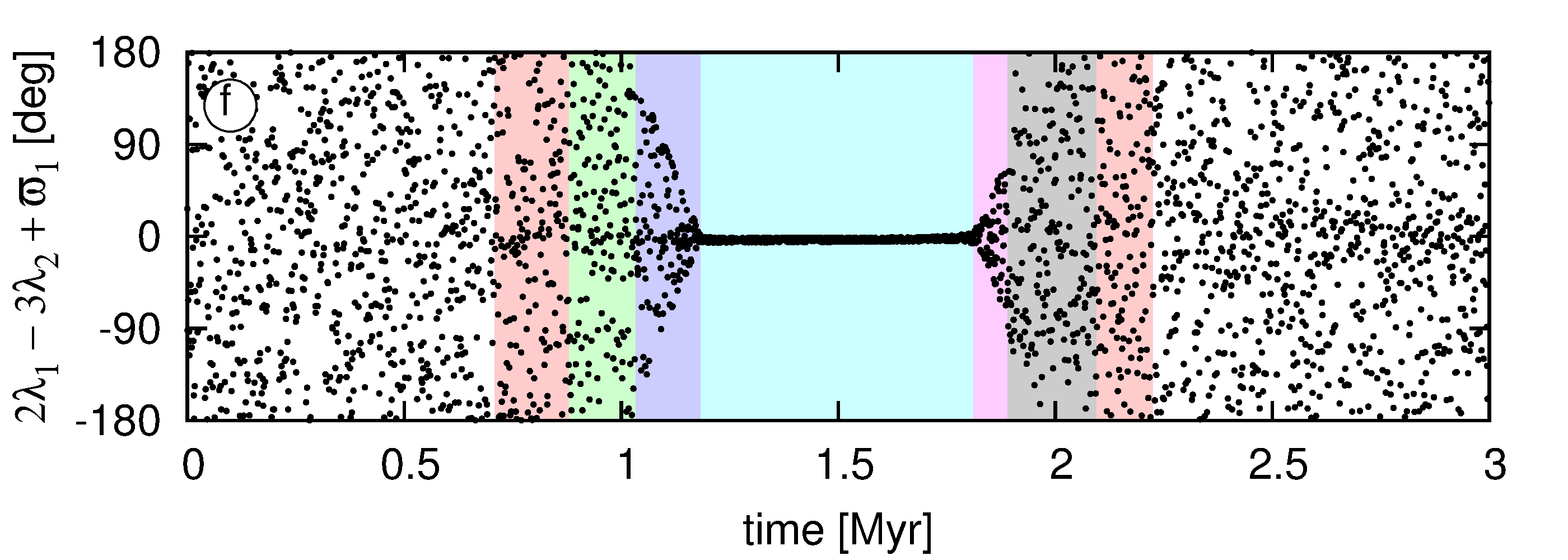}
}
\hbox{
\includegraphics[width=0.49\textwidth]{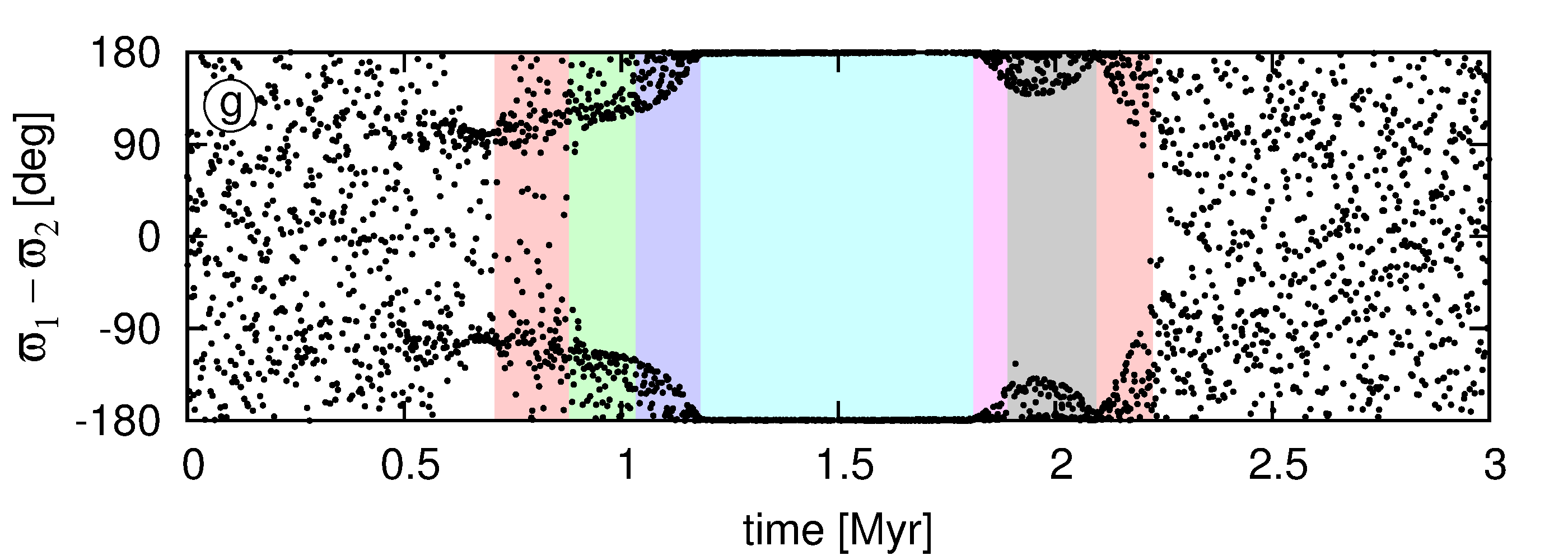}
\includegraphics[width=0.49\textwidth]{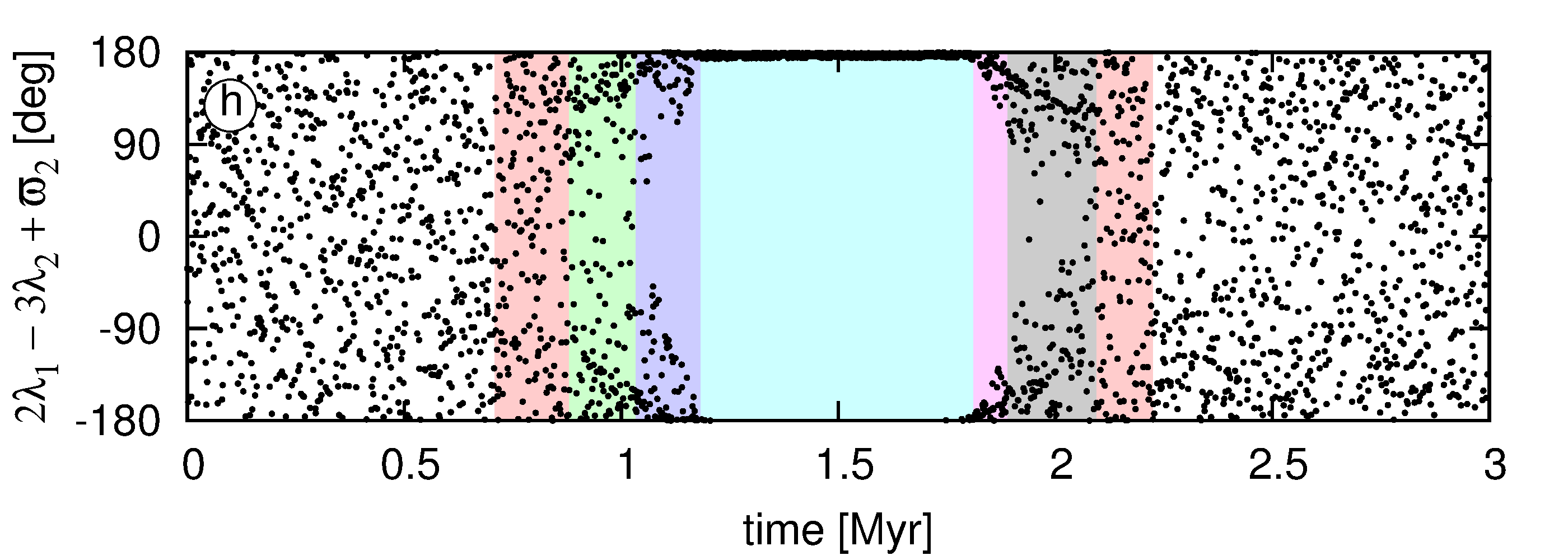}
}
}
}
\caption{Evolution of a chosen initial system of $m_1 = 8\,\mE$, $m_2 = 6\,\mE$, $\log_{10} a_1 = 0.6$, $P_2/P_1 = 2.6$. Initial epoch $t=0$ corresponds to $t=5 \times 10^5\,\yr$ of the disc evolution. Plots of $a$ and $e$ evolution present the results for both planets, the inner one (red) and the outer one (blue). White area bordered with solid black curve in panel~(a) denotes the transition region in a disc between shadowed and irradiated parts. This region corresponds to the outward migration. Remaining panels present the evolution of $P_2/P_1$ and the angles defined at corresponding panel. Characteristic stages of the evolution of the system (different ranges of $t$) are marked with different colours (see the text for details).}
\label{fig:two_planet_evolution_chosen}
\end{figure*}

Before going to the next section, in which the evolution of two-planet systems is studied in more details, one can make an easy observation from the inspection of Fig.~\ref{fig:one_planet_ecc_0}. When two planets are initially located outside the trap and are migrating convergently, they can form a MMR. The convergent migration means that before forming the MMR, the outer planet migrates faster than the inner one ($\tau_1 > \tau_2$). When the planets migrate as a resonant pair, the outer planet migrates slower and the inner planet migrates faster with respect to the non-resonant case. It means that there has to be some torque exchange between the planets. The torque acting on the inner planet, $\Gamma_{1, \idm{res}}$, decreases, and the one for the outer planet, $\Gamma_{2, \idm{res}}$, increases when compared to the non-resonant case, i.e., $\Gamma_{1, \idm{res}} < \Gamma_{1, \idm{non-res}}$ and $\Gamma_{2, \idm{res}} > \Gamma_{2, \idm{non-res}}$. The torque acting on a planet which resides in the red region of the map is positive in absence of the resonant interactions with the outer planet, $\Gamma_{1, \idm{non-res}} > 0$. It is possible that $\Gamma_{1, \idm{res}} < 0$, thus the inner planet can migrate inward and pass through the trap. The same cannot happen for the outer planet, because if $\Gamma_{2, \idm{non-res}} > 0$, also $\Gamma_{2, \idm{res}} > 0$. In such a situation, the outer planet stays at a trap and the system becomes hierarchical ($P_2/P_1 \gg 1$). 
The trap plays a very important role in forming the architecture of two-planet systems. In configurations with more than two planets, a ''convoy'' of planets can lead to a situation in which all the planets, apart from the outermost one, pass through the trap and reach inner regions of the disc. We will study systems with larger number of planets in the next paper.

Another important feature of the $\tau_a-$maps is that 
at a given epoch $\tau_a$ has minima and maxima along $\log_{10} a$. The divergent migration occurs when $\tau_{a,1} < \tau_{a,2}$ (if both $\tau$s are positive). It means that there are orbits for which a system, which initially resides in MMR, starts to diverge out of the resonance. It happens when the inner planet is close to the local minimum of $\tau_a$. Although such a simple reasoning is valid for a case of equal masses $m_1 = m_2$, it shows that regions of divergent migration should exist in the disc.

\section{Evolution of two--planet systems}

If $\tau_{a,1}$ and $\tau_{a,2}$ are the time-scales of the migration of the inner and the outer planets in a pair, respectively, the time-scale of the period ratio variation reads as follows:
\begin{equation}
\tau_X \equiv - \frac{X}{\dot{X}}, \quad
X \equiv \frac{P_2}{P_1}, \quad
\tau_X = \frac{2}{3} \, \frac{\tau_{a,1} \, \tau_{a,2}}{\tau_{a,1} - \tau_{a,2}}.
\end{equation}
If both $\tau_{a,1}$ and $\tau_{a,2}$ are positive and $\tau_{a,1} > \tau_{a,2}$, than $\tau_X > 0$ and the period ratio decreases. If $\tau_{a,1} < \tau_{a,2}$, the migration is divergent. Similarly to the migration maps presented in Fig.~\ref{fig:one_planet_ecc_0}, one can make maps for $\tau_X$. 
Figures~\ref{fig:two_planet_7_7_0_0}a,b and~c present $\tau_X$-maps on $(P_2/P_1, \log_{10} a_1)-$plane computed for $t = 5 \times 10^5\,\yr$ and three different pairs of masses $m_1 = m_2 = 7\,\mE$ (panel a), $m_1 = 6\,\mE, m_2 = 8\,\mE$ (b) and $m_1 = 8\,\mE, m_2 = 6\,\mE$ (c). If a two-planet system is located in a green region, $P_2/P_1$ will decrease, while if the system is located in red regions, the period ratio will increase. As colours give only the information on the $P_2/P_1$ evolution, the arrows are over-plotted on the maps to complete this information. If the arrows point down/up, the migration of the inner planet is inward/outward. Directions left/right correspond to convergent/divergent migration.

The view of the $(P_2/P_1, \log_{10} a_1)-$plane depends on the planets masses. When the inner planet is less massive than the outer one, larger fraction of the plane corresponds to the convergent migration (see Fig.~\ref{fig:two_planet_7_7_0_0}b). When $m_1 > m_2$, most of the plane corresponds to the divergent migration (see Fig.~\ref{fig:two_planet_7_7_0_0}c). One might expect that for $m_1 < m_2$ MMRs form more frequently than for $m_1 > m_2$. 

As the disc evolves, the view of the $\tau_X-$maps changes. Fig~\ref{fig:two_planet_7_7_0_0}d presents the map for $m_1=8\,\mE$, $m_2=6\,\mE$ at $t = 1.5\times10^6\,\yr$. Also the position of the system in the plane changes (compare positions of an example system plotted with white symbols at panels c and d). It means that during the evolution, the system may reside in a convergent/divergent region some limited time and the period ratio may increase or decrease depending on the evolution stage. The final configuration depends in a complex way one the initial orbits, planets' masses and disc evolution time scale.

\subsection{An example of the evolution of a two-planet system}

Figure~\ref{fig:two_planet_evolution_chosen} presents the evolution of an example system. Both planets of masses $m_1 = 8\,\mE$ and $m_2 = 6\,\mE$ start the migration outside the trap (see Fig.~\ref{fig:two_planet_evolution_chosen}a, the white region bordered with black solid curve denotes the shadowed/irradiated disc transition region, see also the position of the system plotted with white symbol in Fig.~\ref{fig:two_planet_7_7_0_0}c) at the epoch $0.5~$Myr of the disc evolution. Ranges of time, corresponding to characteristic behaviour of the system, are colour-coded. 

\begin{figure}
\centerline{
\includegraphics[width=0.49\textwidth]{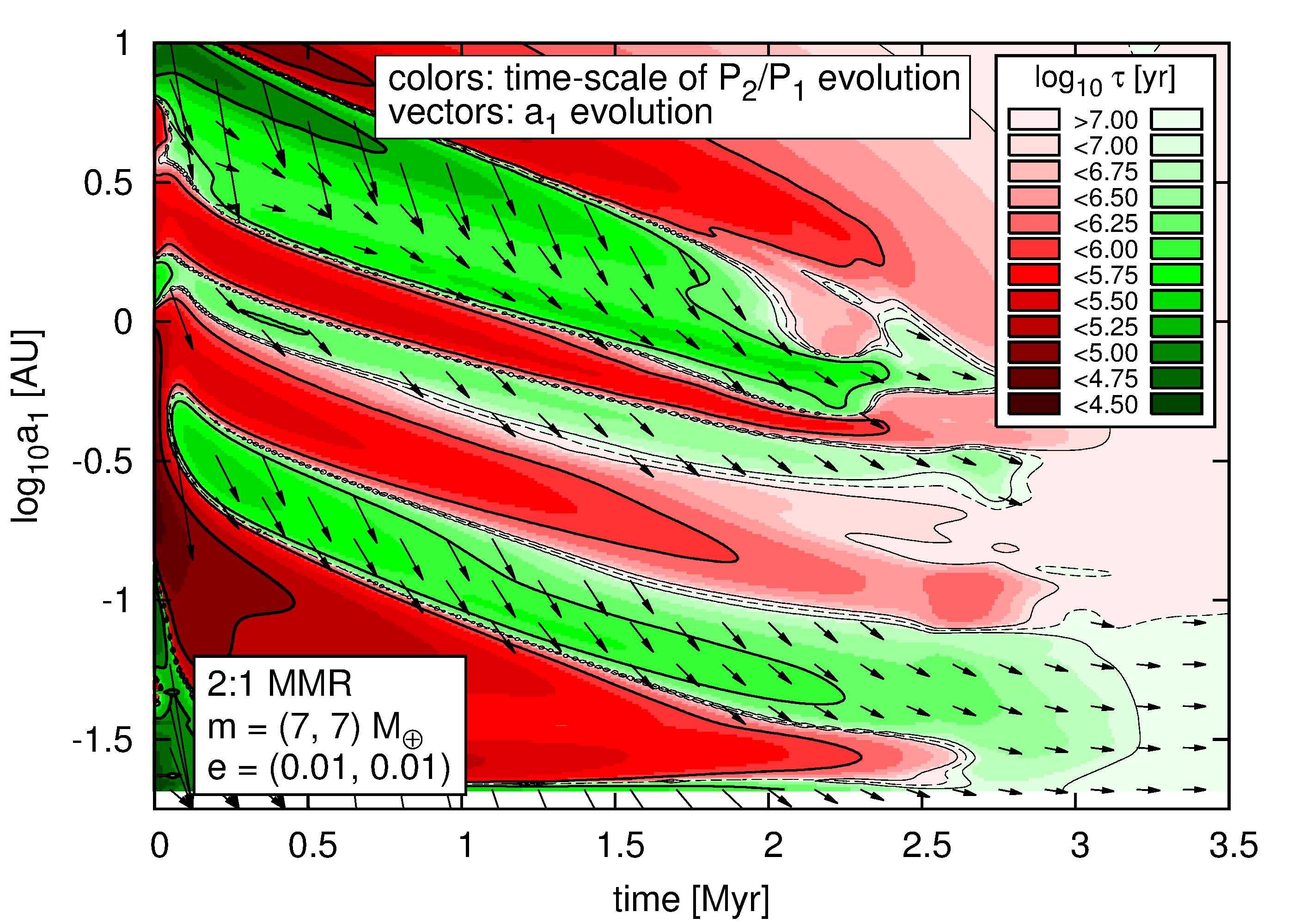}
}
\caption{Time-scale $\tau_X$ (colours) and a vector filed for the inner planet, which is involved in 2:1~MMR with the outer planet. $m_1 = m_2 = 7\,\mE$.}
\label{fig:21MMR}
\end{figure}

Initially the planets migrate convergently, the period ratio decreases from $2.6$ down to $\sim 1.65$ (Fig.~\ref{fig:two_planet_evolution_chosen}c). The convergent migration is very fast, thus the system crosses 2:1~MMR. It does not reach 3:2~MMR, because before that it moves into the region of the divergent migration. The period ratio increases approaching $2$. The first critical angle of 2:1~MMR $\phi_{2:1}^{(1)} \equiv \lambda_1 - 2\,\lambda_2 + \varpi_1$ (Fig.~\ref{fig:two_planet_evolution_chosen}b) librates when $P_2/P_1$ is sufficiently close to $2$. This period of the evolution is coloured with light red.
At $t \sim 0.9~$Myr the inner planet reaches the trap, while the outer planet keeps migrating inward. The result is that $P_2/P_1$ decreases again and the system reaches 3:2~MMR. Before that, for $t \in (\sim 0.9, \sim 1)~$Myr, when $P_2/P_1$ is still close to $2$, both resonant angles of 2:1~MMR librate (the second angle is defined as $\phi_{2:1}^{(2)} \equiv \lambda_1 - 2\,\lambda_2 + \varpi_2$, see Fig.~\ref{fig:two_planet_evolution_chosen}d). This fragment of the evolution is coloured with light green. When both critical angles librate, also the difference between the longitudes of pericenter, $\Delta\varpi \equiv \varpi_1 - \varpi_2$, librates (Fig.~\ref{fig:two_planet_evolution_chosen}g).
The period ratio keeps decreasing and when $P_2/P_1 \sim 1.7$ the critical angles of 2:1~MMR start to rotate and the angles of 3:2~MMR ($\phi_{3:2}^{(1)} \equiv 2\,\lambda_1 - 3\,\lambda_2 + \varpi_1$ and $\phi_{3:2}^{(2)} \equiv 2\,\lambda_1 - 3\,\lambda_2 + \varpi_2$, see Fig.~\ref{fig:two_planet_evolution_chosen}f,h) start to librate. This fragment of $t$-axis, which corresponds to trapping the system into 3:2~MMR is coloured with light violet. At $t \sim 1.1~$Myr $P_2/P_1$ approaches $1.5$ and the amplitudes of the librations of both resonant angles are very small. The eccentricities reach their equilibrium values $\lesssim 0.01$. The planets migrate as a resonant pair for $\sim 0.5~$Myr (light blue colour). The inner planet passes through the outward migration region at $t \sim 1.7~$Myr. The outer planet reaches the trap and migrate at the viscous time-scale. The inner planet migrates faster, thus the system goes out of 3:2~MMR. The resonant angles of 3:2~MMR librate until $P_2/P_1 \lesssim 1.7$, after which the angles of 2:1~MMR start to librate again. When $P_2/P_1$ passes through $2$, the behaviour of the angles changes. The first angle librates around $0$ instead of $180~$degrees, and the second angle librates around $180$ instead of $0~$degrees. The second angle librates only when $P_2/P_1 \lesssim 2.12$. 

The final period ratio is $\sim 2.4$ and only $\phi_{2:1}^{(1)}$ oscillates. Both eccentricities are very small $\sim 10^{-4}$. As shown in \citep{Batygin2013}, for very low $e$ the frequency of the apsidal motion can be high enough to make the critical angle oscillate even for $P_2/P_1$ significantly distant from $(p+1)/p$. \cite{Papaloizou2011} also studied resonant behaviour of a system far from the nominal position of the resonance. As the system moves towards higher $P_2/P_1$, its eccentricities are very small, but the resonant angles keep librating.

\subsection{Conditions for keeping a system in MMR}

Once a system is locked in MMR, one can ask at what radii and what evolution stage of the disc this configuration can be preserved. As we illustrated in the previous section, planets can remain in the region of the convergent migration only for some limited time, depending on the time-scales of their migration and the disc evolution. Figure~\ref{fig:21MMR} shows $\tau_X$-maps for 2:1~MMR for $m_1 = m_2 = 7\,\mE$. For a given $a_1$, a value of $a_2$ is such that $P_2/P_1=2$. If initially resonant configuration is placed in the red region, the system will move out of the resonance. If the system is placed in the green region, its period ratio will be constant (although the green colour means that $P_2/P_1$ decreases, it holds true only for a non-resonant case).

\begin{figure*}
\centerline{
\vbox{
\hbox{
\includegraphics[width=0.49\textwidth]{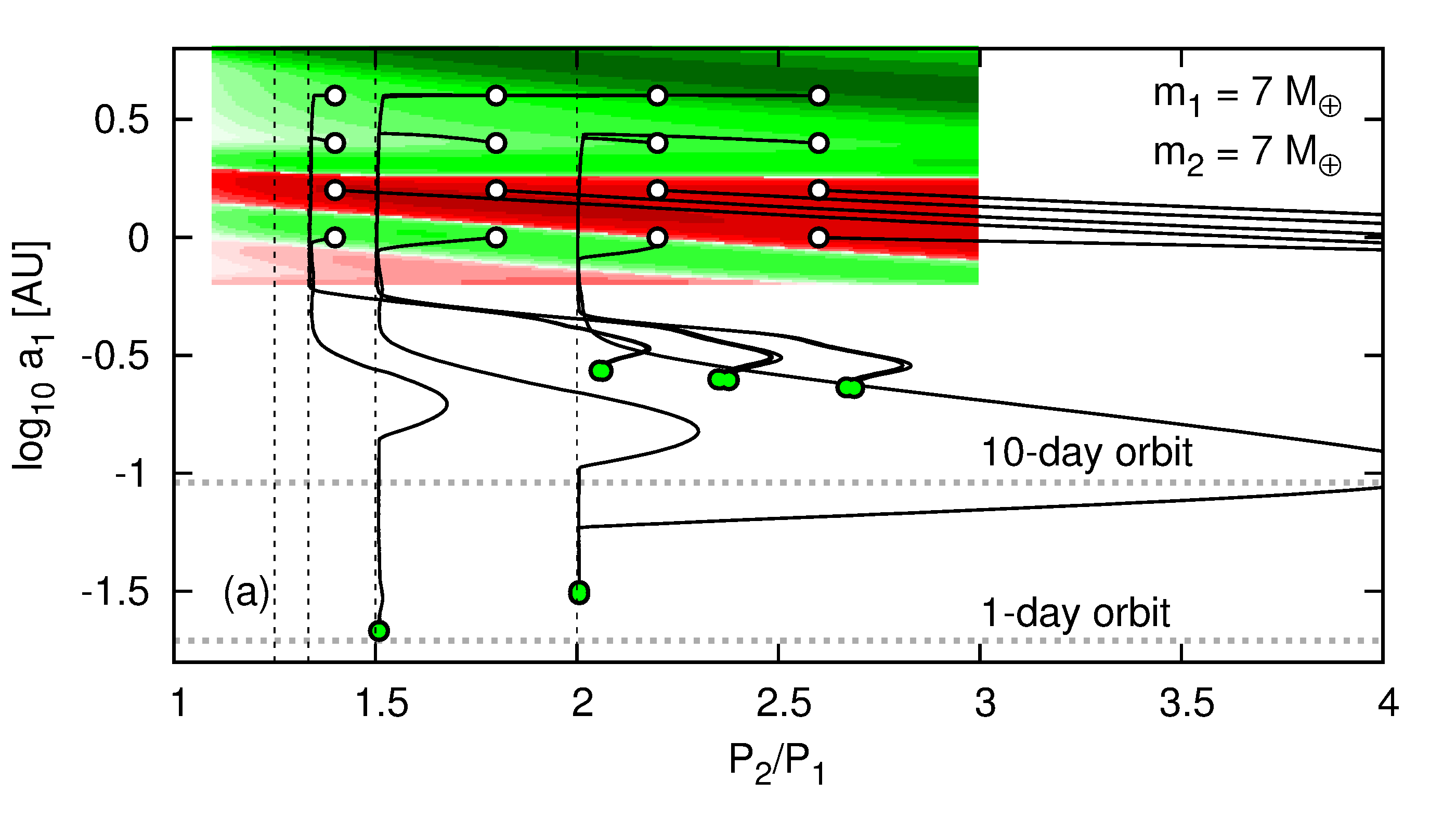}
\includegraphics[width=0.49\textwidth]{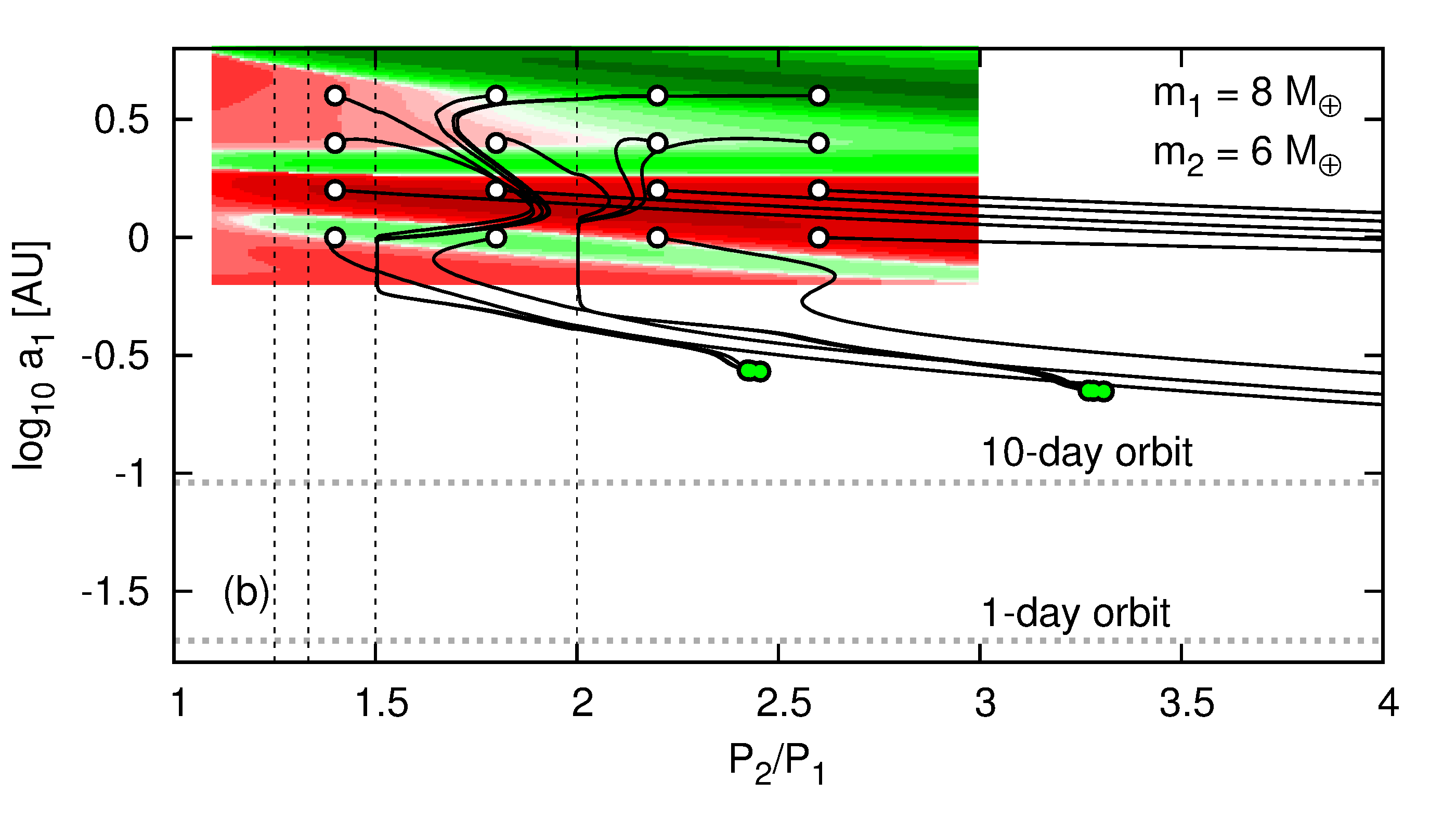}
}
}
}
\caption{The evolution of $16$ initial systems (white dots) plotted with black solid curves. Final configurations are marked with green dots. Planets' masses are given in top right-hand corner of each panel. The $\tau_X-$maps for the beginning of the simulation is colour-coded. Vertical dashed lines denote MMRs.}
\label{fig:two_planet_evolution}
\end{figure*}

If the system is non-resonant, the evolution of the semi-major axes of the planets are given by formulae:
\begin{equation}
\dot{a}_1^{\idm{(non-res)}} = \frac{2\,a_1\,\Gamma_1}{L_{1,0}}, \quad
\dot{a}_2^{\idm{(non-res)}} = \frac{2\,a_2\,\Gamma_2}{L_{2,0}},
\end{equation}
where $\Gamma_1$ is the torque acting on the inner planet and $\Gamma_2$ is the torque acting on the outer planet. The angular momenta of the planets moving in circular orbits are given by $L_{1,0}$ and $L_{2,0}$ for the inner and the outer planet, respectively.
When the system is locked in MMR, the ratio of the semi-major axes is constant. Another constrain is that the sum of the torques acting on the planets is the same in both resonant and non-resonant situations. Therefore we have the following formulae for $\dot{a}$ in the resonant case:
\begin{equation}
\dot{a}_1^{\idm{(res)}} = \frac{2\,a_1\left(\Gamma_1 + \Gamma_2\right)}{L_{1,0} + L_{2,0}}, \quad
\dot{a}_2^{\idm{(res)}} = \frac{2\,a_2\left(\Gamma_1 + \Gamma_2\right)}{L_{1,0} + L_{2,0}}.
\end{equation}
The vector field for $\dot{a}_1$ (shown in Fig.~\ref{fig:21MMR}) corresponds to the resonant case. As we can see, 
the inner planet migrates faster than the disc evolves. Sooner or later an initially resonant configuration moves into the region of the divergent migration. At $t \sim 2~$Myr the orbits of $\log_{10} a_1 [\au] \sim -0.4, \in (-1, -0.5), \sim -1.5$ correspond to the divergent migration, and systems located there cannot remain in 2:1~MMR. At $t \gtrsim 3~$Myr the resonance can be sustained only for $r \lesssim 0.1\,\au$. When the inner planet is less massive than the outer one most of the plane corresponds to the convergent migration and the 2:1~MMR can be formed and preserved for almost all radii $\log_{10} a_1 [\au] \lesssim -0.5$. When $m_1 > m_2$, most of the $\tau_X-$map corresponds to the divergent migration. In such a case, a resonant system is unlikely to form due to migration.

\subsection{Migration for different masses and initial orbits}

As we showed above, the final orbital configuration depends on the masses of the planets as well as on the initial orbits. Figure~\ref{fig:two_planet_evolution} shows trajectories of systems starting from different initial positions at the $(P_2/P_1, \log_{10} a_1)-$plane. 
Panel~(a) shows the results obtained for $m_1 = m_2 = 7\,\mE$. Initial positions of sixteen systems are marked with white symbols over-plotted at $\tau_X-$map computed for the initial epoch. Eight of the systems (group~$1$) initially reside above the red region, another five (group~$2$) are located within this region, while three systems (group~$3$) below the region. The systems from the first group initially tends towards 2:1, 3:2 or 4:3~MMR. The planets migrate slightly faster than the disc evolves. They reach the red region, i.e., the inner planet passes through the outward migration region and migrate further, while the outer planet is halted at the trap (see the discussion in Section~4). The systems start to diverge from the resonances. After the trap vanishes at $t \sim 2.3~$Myrs (see  Fig.~\ref{fig:one_planet_ecc_0}), the systems are again located in the convergent migration zone. However, because the disc is of low mass and shortly after that moment it disperses, the systems do not have enough time to reach 2:1~MMR. The final $P_2/P_1$ are between $2$ and $3$.

The systems from the second group evolve towards large $P_2/P_1$. In this case, the inner planet in each system is initially below the trap and migrates inward, while the outer planet is stopped at the trap. Until the trap disappears letting the outer planet to migrate inward, the inner one is already very close to the star. These systems end up as hierarchical configurations. Their final period ratios are larger than $4$, thus they are not shown in the plot.

The systems from the third group reach the resonances (2:1, 3:2 or 4:3) and because they migrate faster than the disc evolves, they avoid the trap. However, at some moment of the evolution, the systems reach the second region of divergent migration (see Fig.~\ref{fig:21MMR}, the red band initially located at $a_1 \sim 1\,\au$ and moving inward down to $a_1 \sim 0.1\,\au$). For some time (when a given system resides in the divergent migration region) their period ratios increase. For instance the system initially located at $(P_2/P_1, \log_{10} a_1 [\au]) = (1.4, 0)$ goes out of 4:3~MMR, and pass through 3:2~MMR. After the systems leave the region of divergent migration, they go towards MMRs (not necessarily the ones they left). After being locked in MMRs, the systems migrate inward as resonant pairs until they reach the third region of divergent migration (see Fig.~\ref{fig:21MMR}, at $\log_{10} a_1 [\au] \sim -1.5$ at $t \gtrsim 2~$Myr). They again diverge slightly out of the resonances and return. The final configurations are resonant and short-period ($P_1$ in each pair ranges from one to a few days). 

Different behaviour can be observed in Fig.~\ref{fig:two_planet_evolution}b. When $m_1 > m_2$ most of the $\tau_X-$map is red (see Fig.~\ref{fig:two_planet_7_7_0_0}c,d) and none of the systems ends up as resonant or close-to-resonant configuration. For a reference, the evolution of the system initially located at $\log_{10}a_1[\au] = 0.6$ and $P_2/P_1 = 2.6$ was shown in Fig.~\ref{fig:two_planet_evolution_chosen}.

\begin{figure*}
\centerline{
\vbox{
\hbox{
\includegraphics[width=0.32\textwidth]{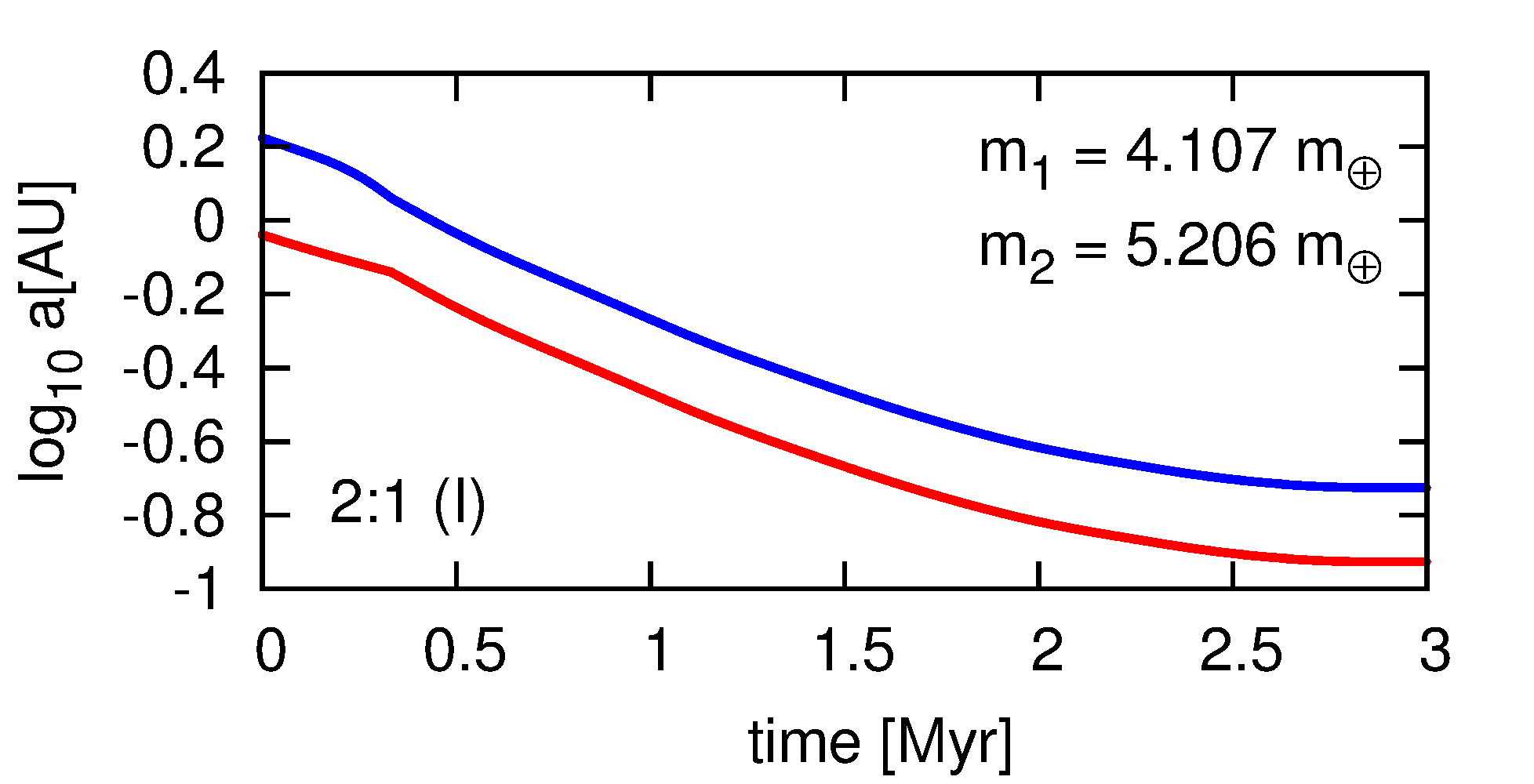}
\includegraphics[width=0.32\textwidth]{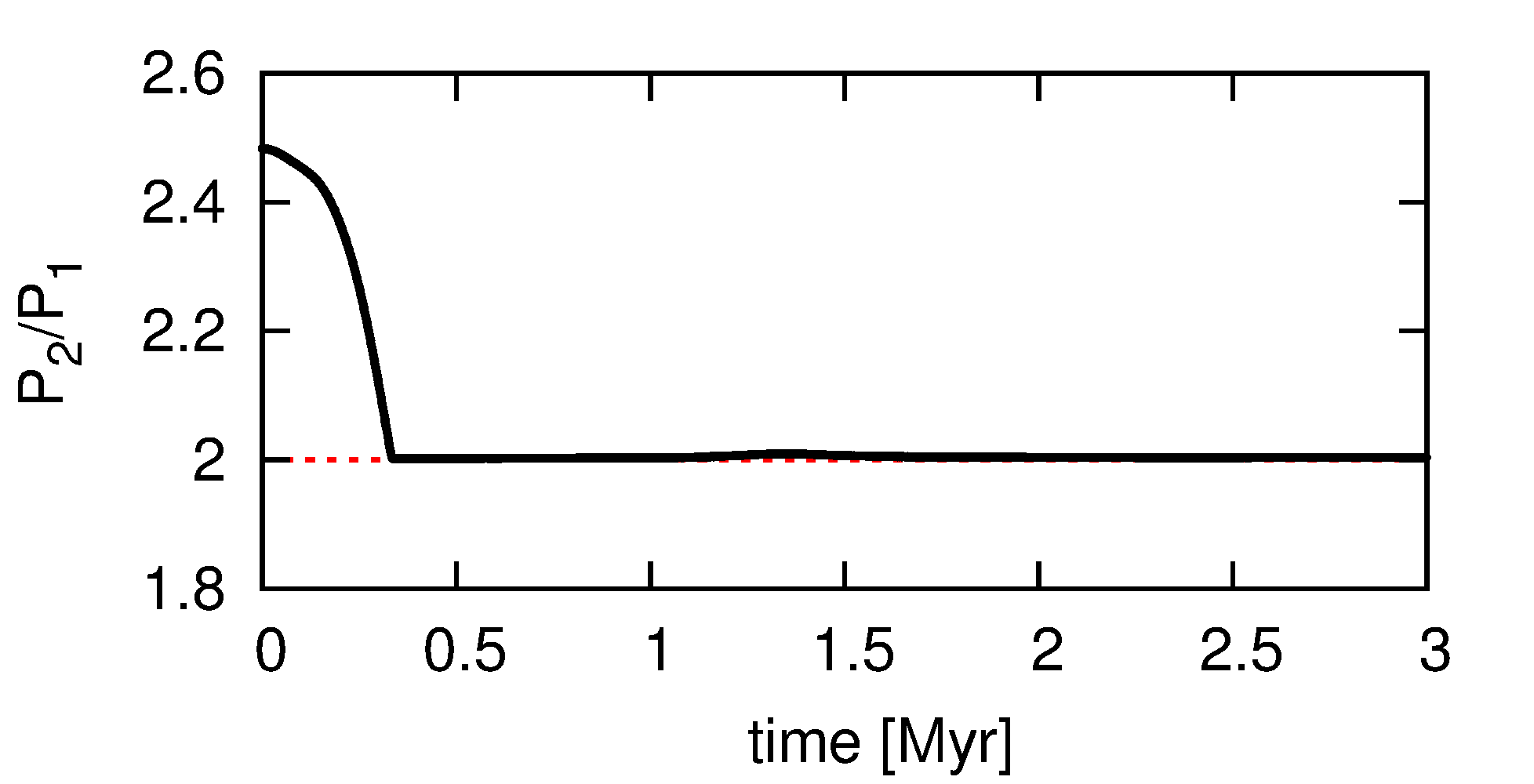}
\includegraphics[width=0.32\textwidth]{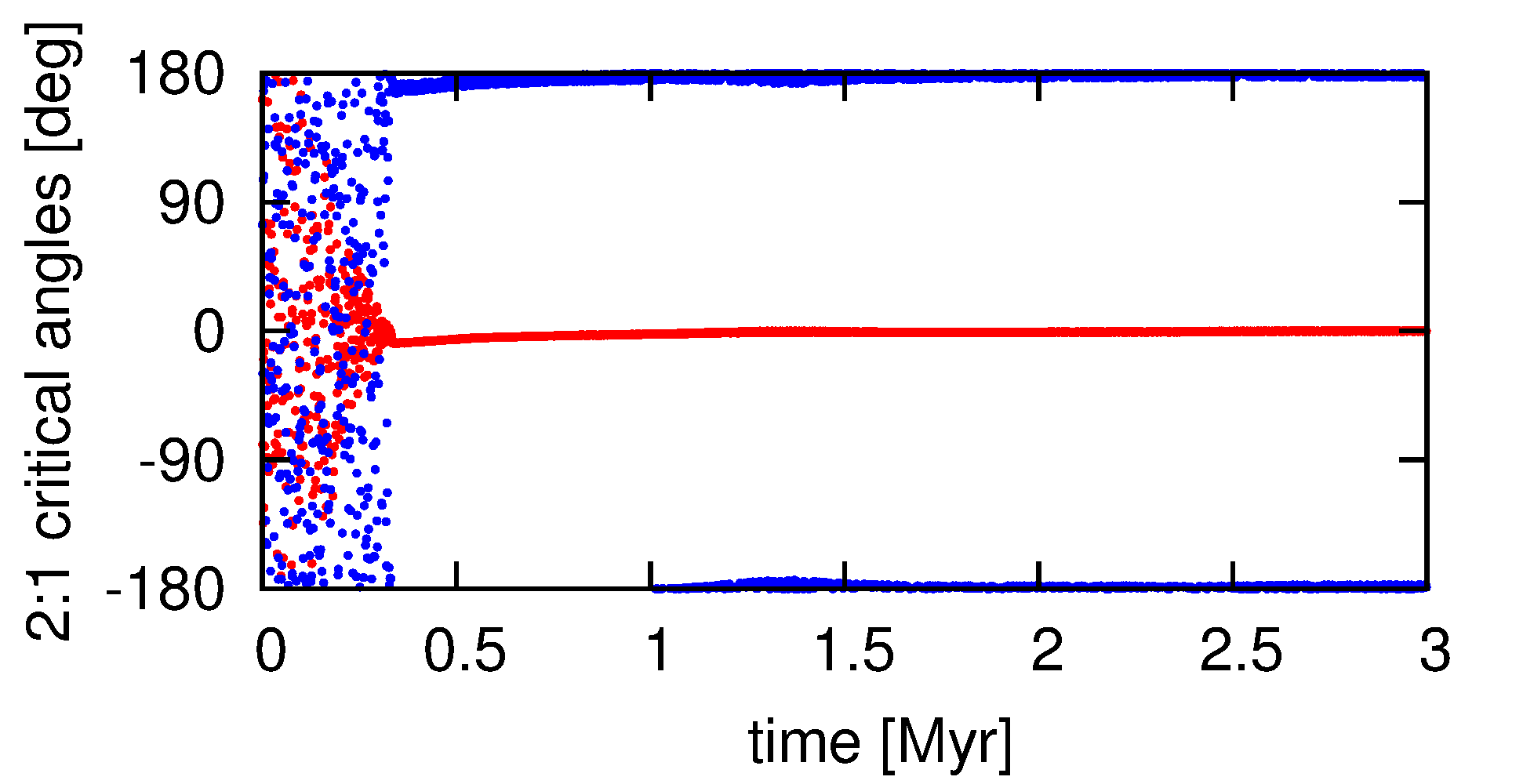}
}
\hbox{
\includegraphics[width=0.32\textwidth]{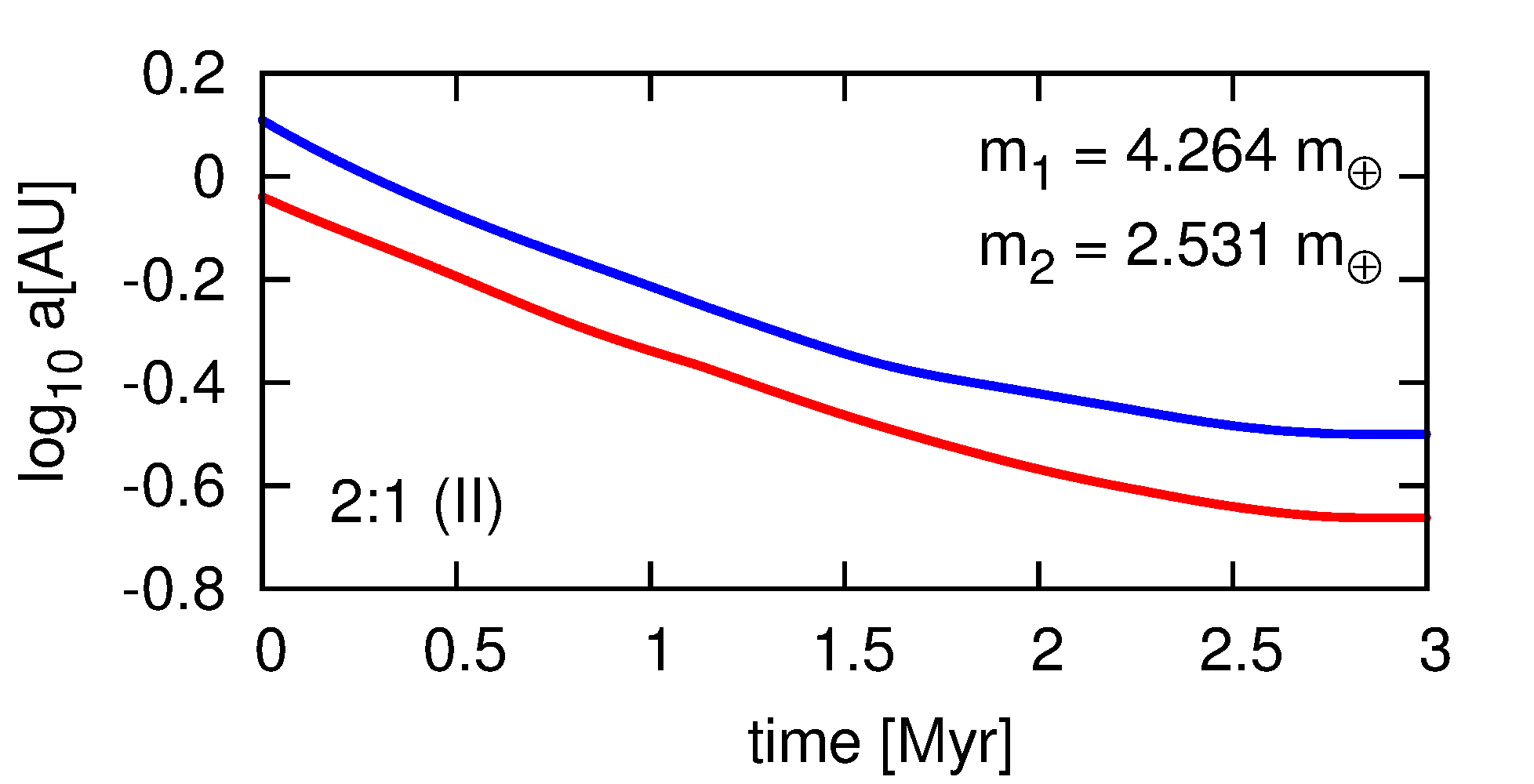}
\includegraphics[width=0.32\textwidth]{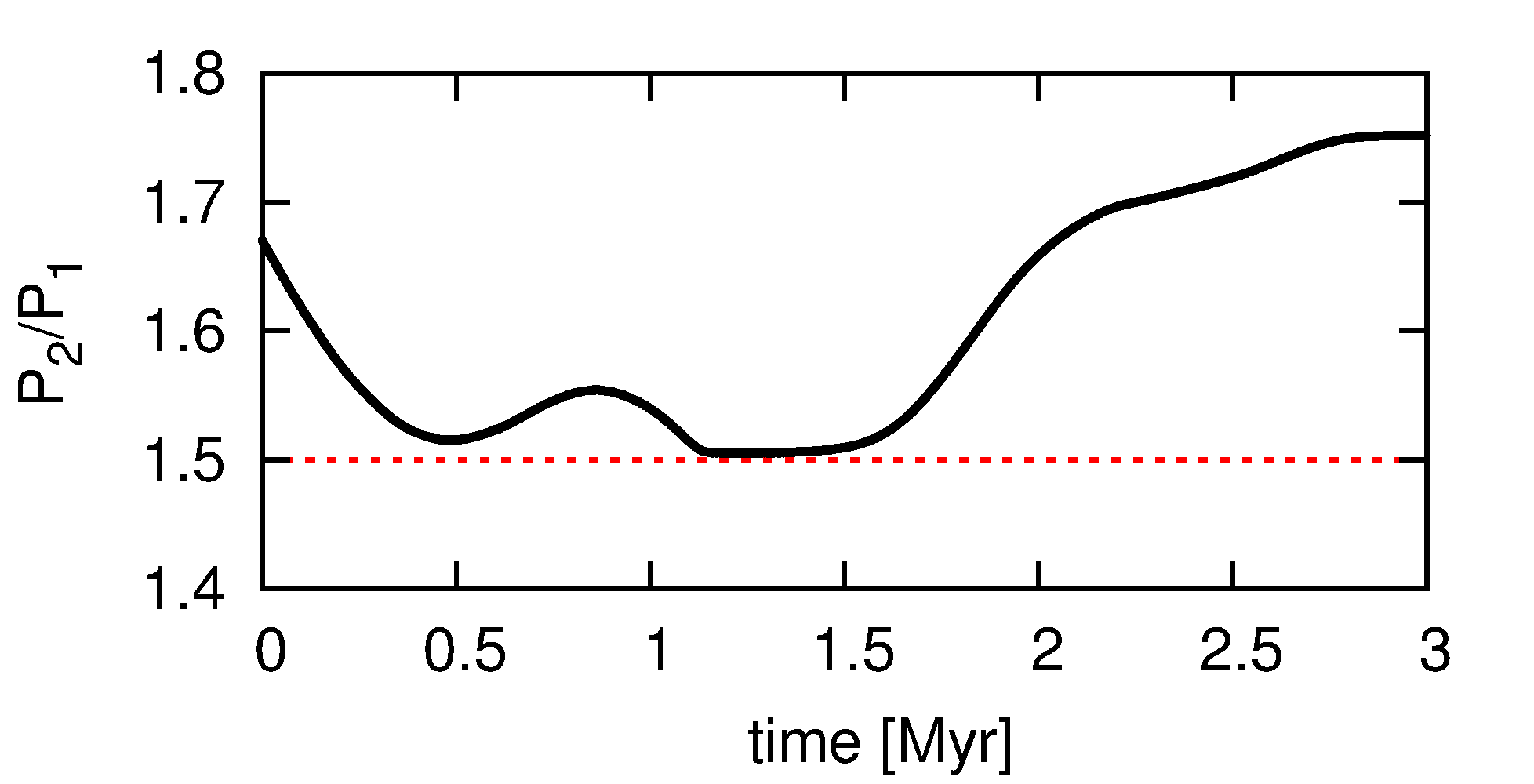}
\includegraphics[width=0.32\textwidth]{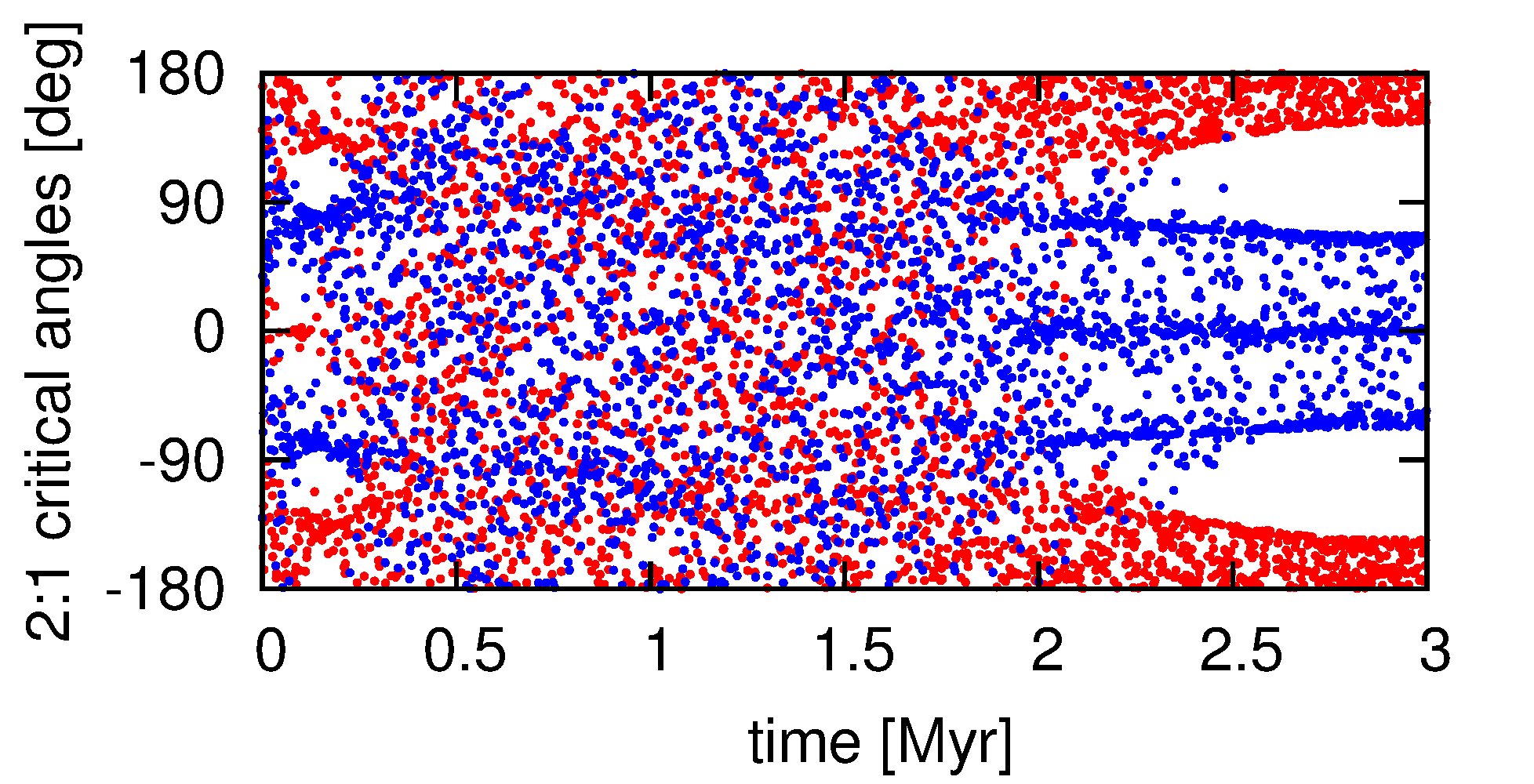}
}
\hbox{
\includegraphics[width=0.32\textwidth]{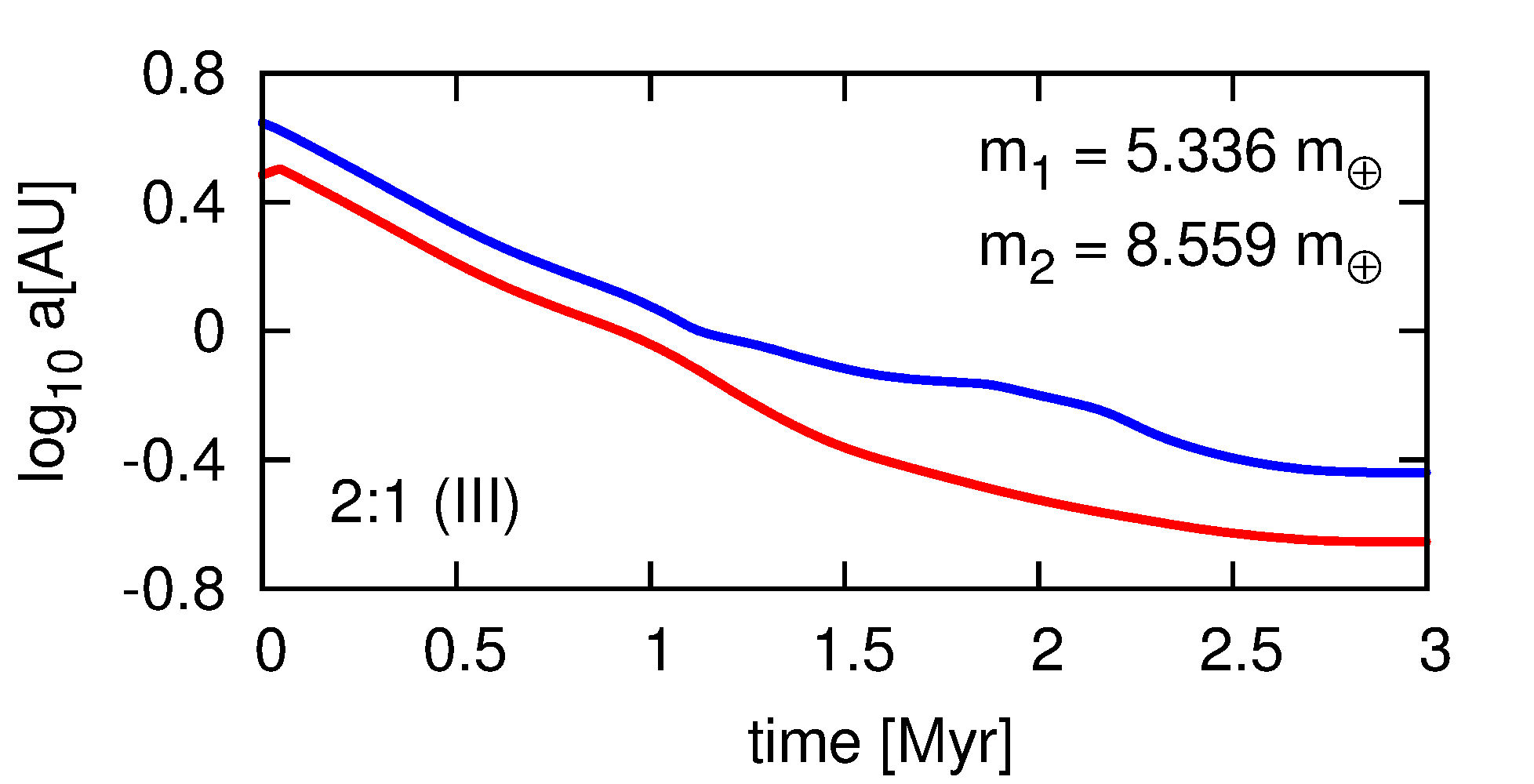}
\includegraphics[width=0.32\textwidth]{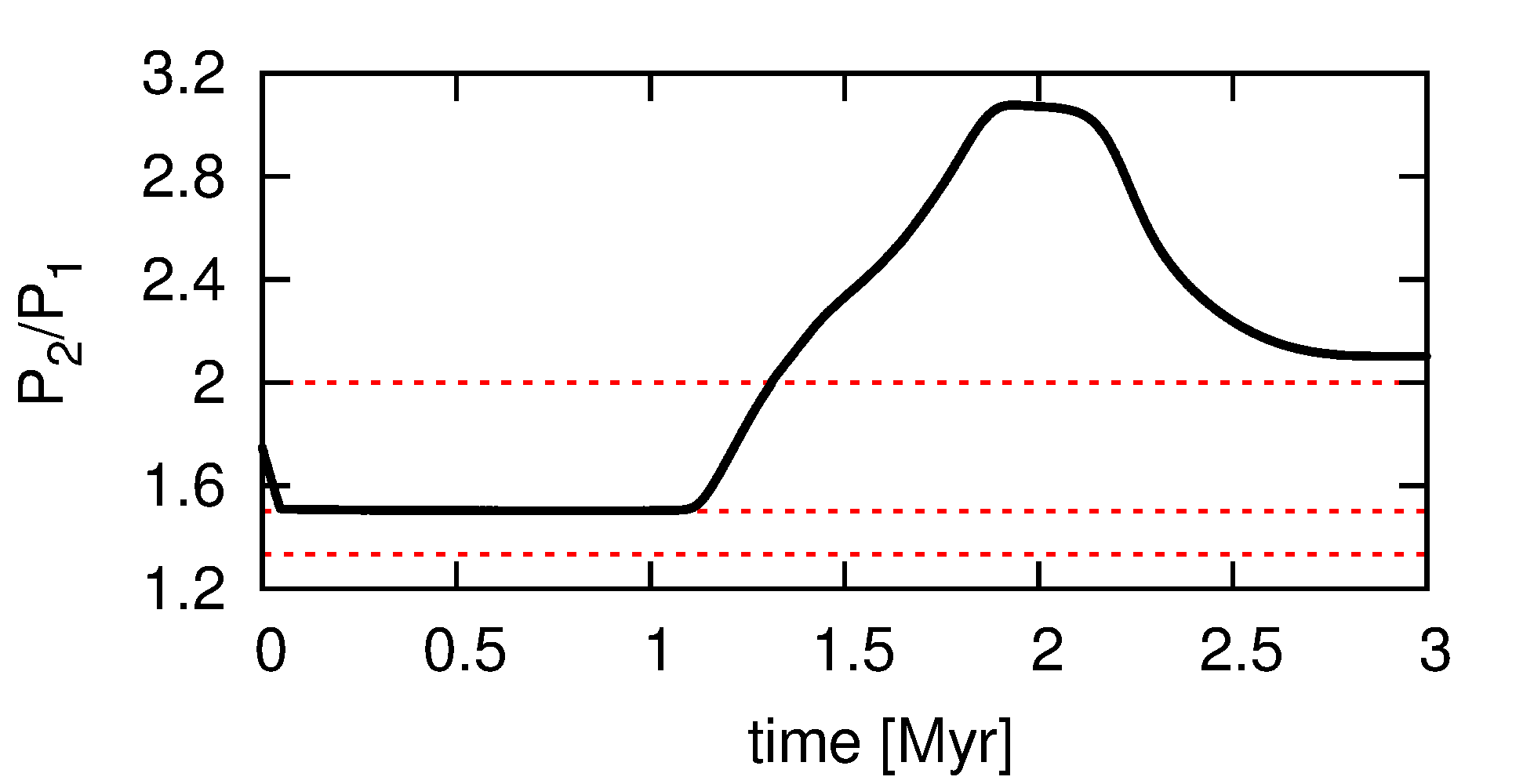}
\includegraphics[width=0.32\textwidth]{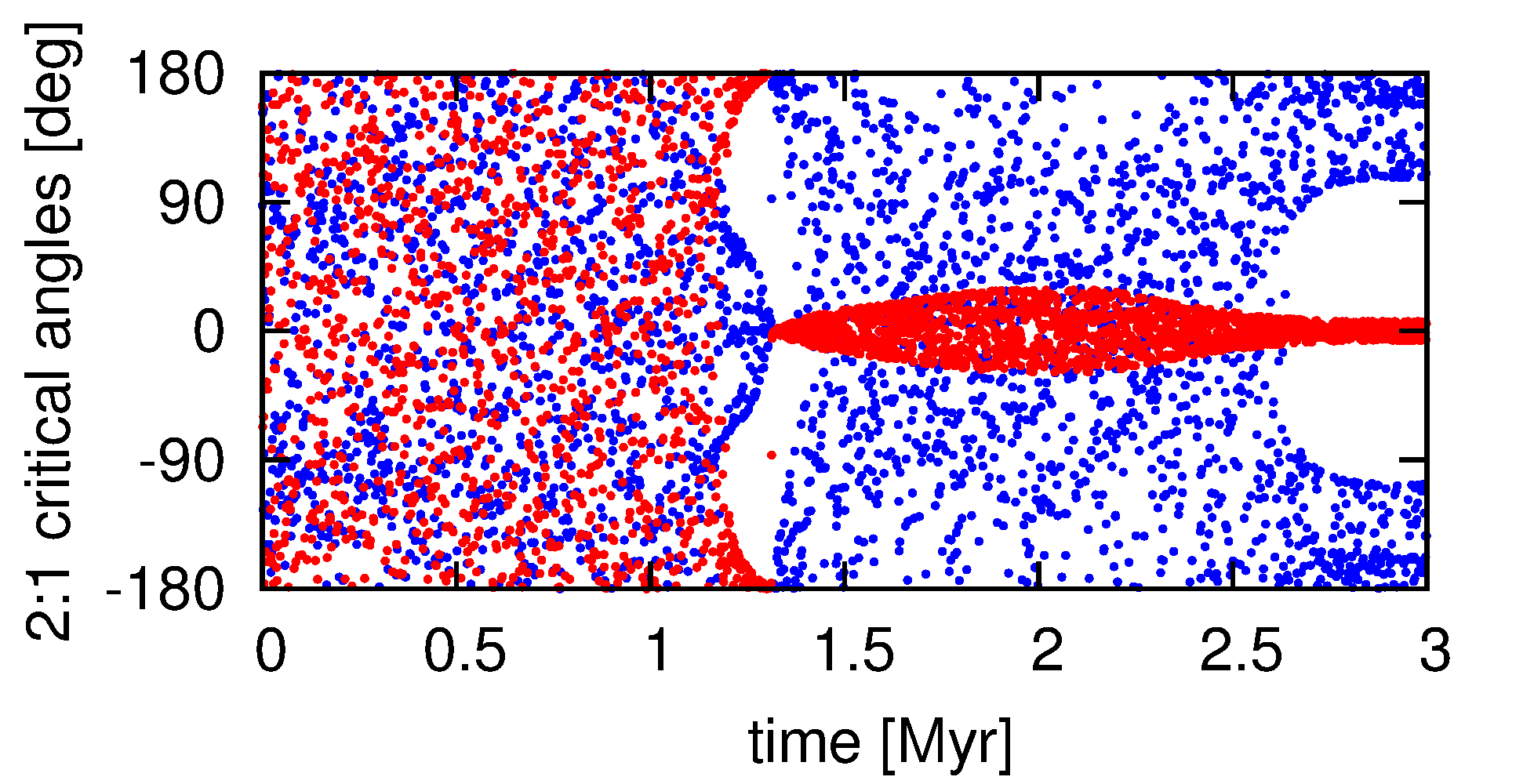}
}
\hbox{
\includegraphics[width=0.32\textwidth]{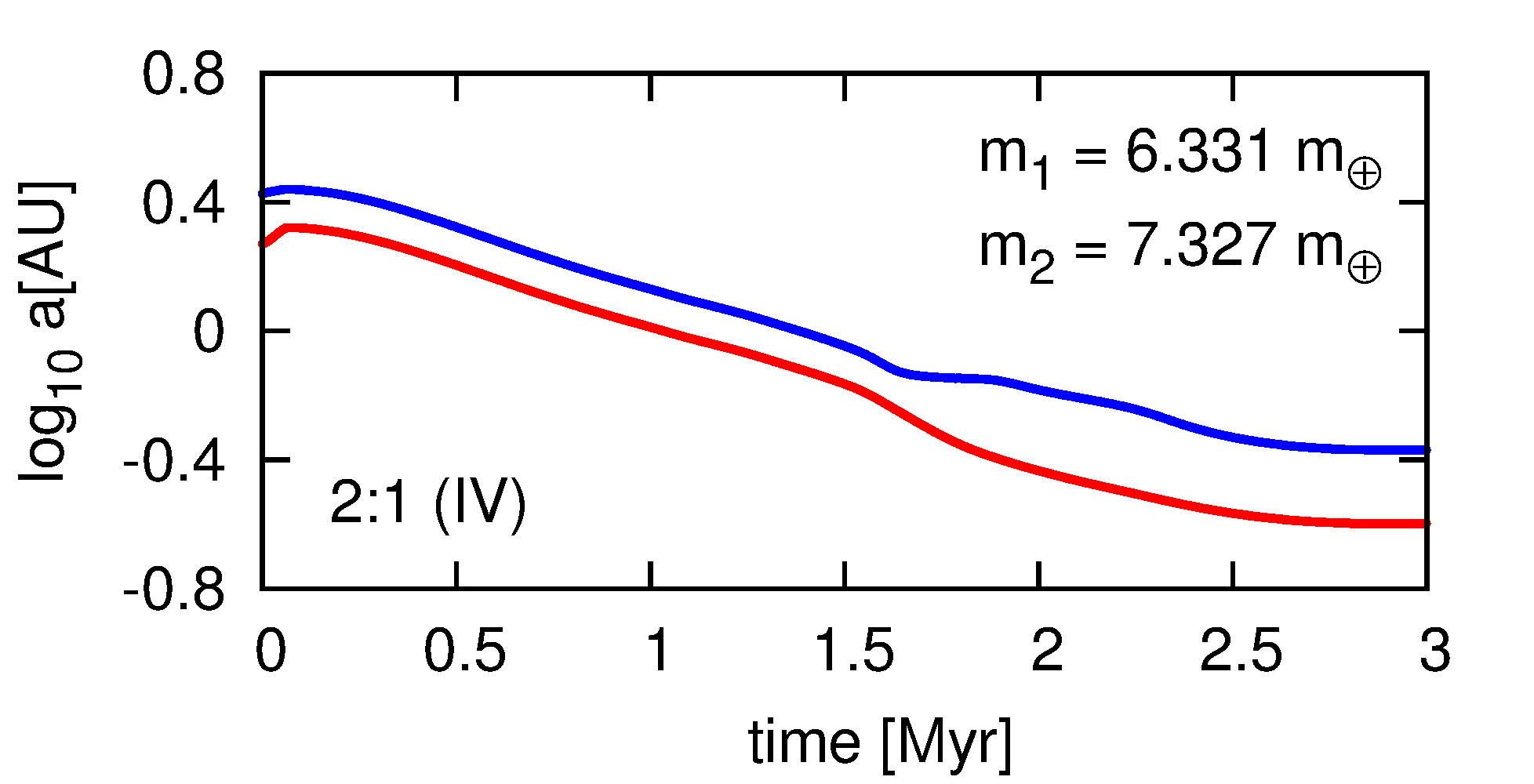}
\includegraphics[width=0.32\textwidth]{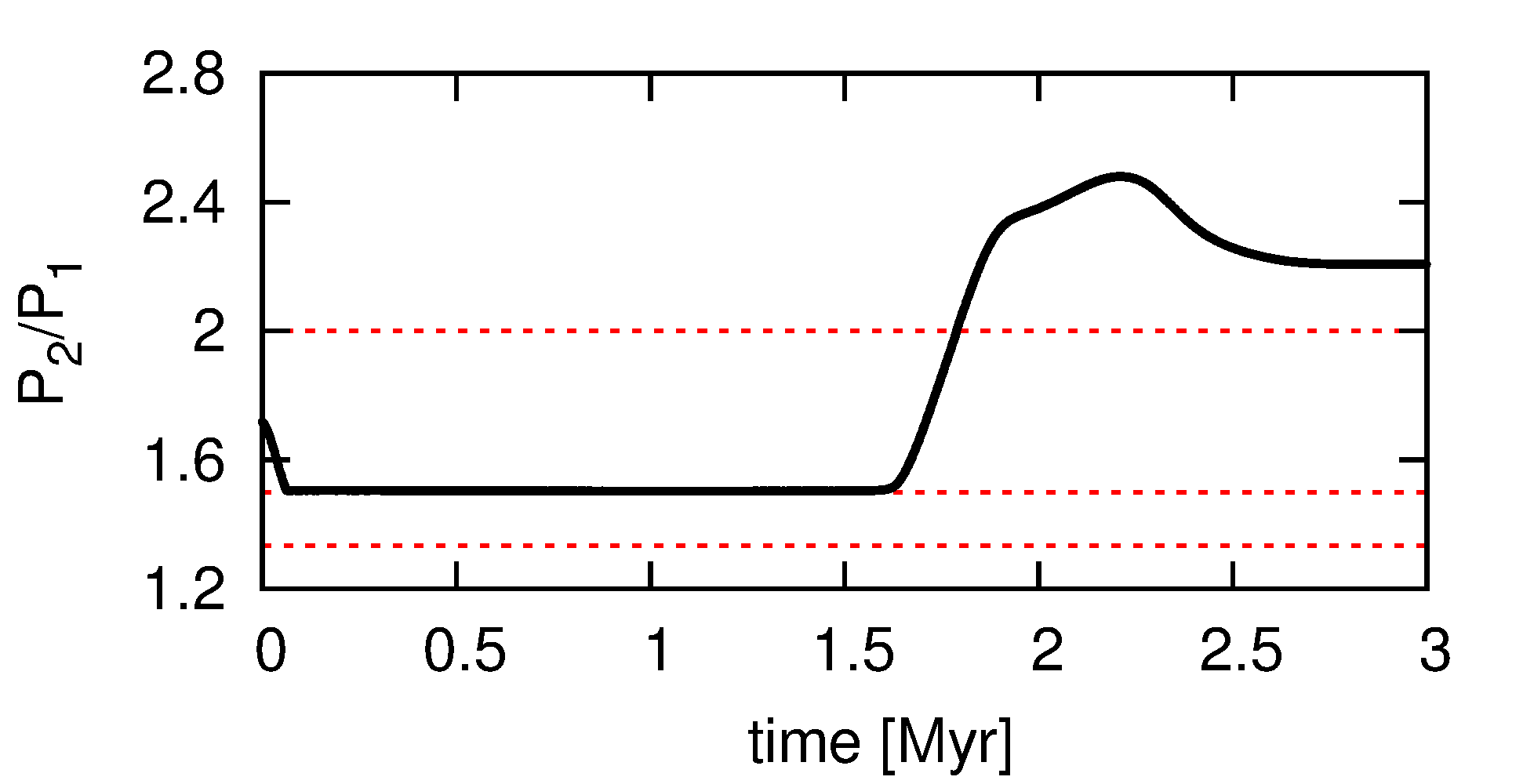}
\includegraphics[width=0.32\textwidth]{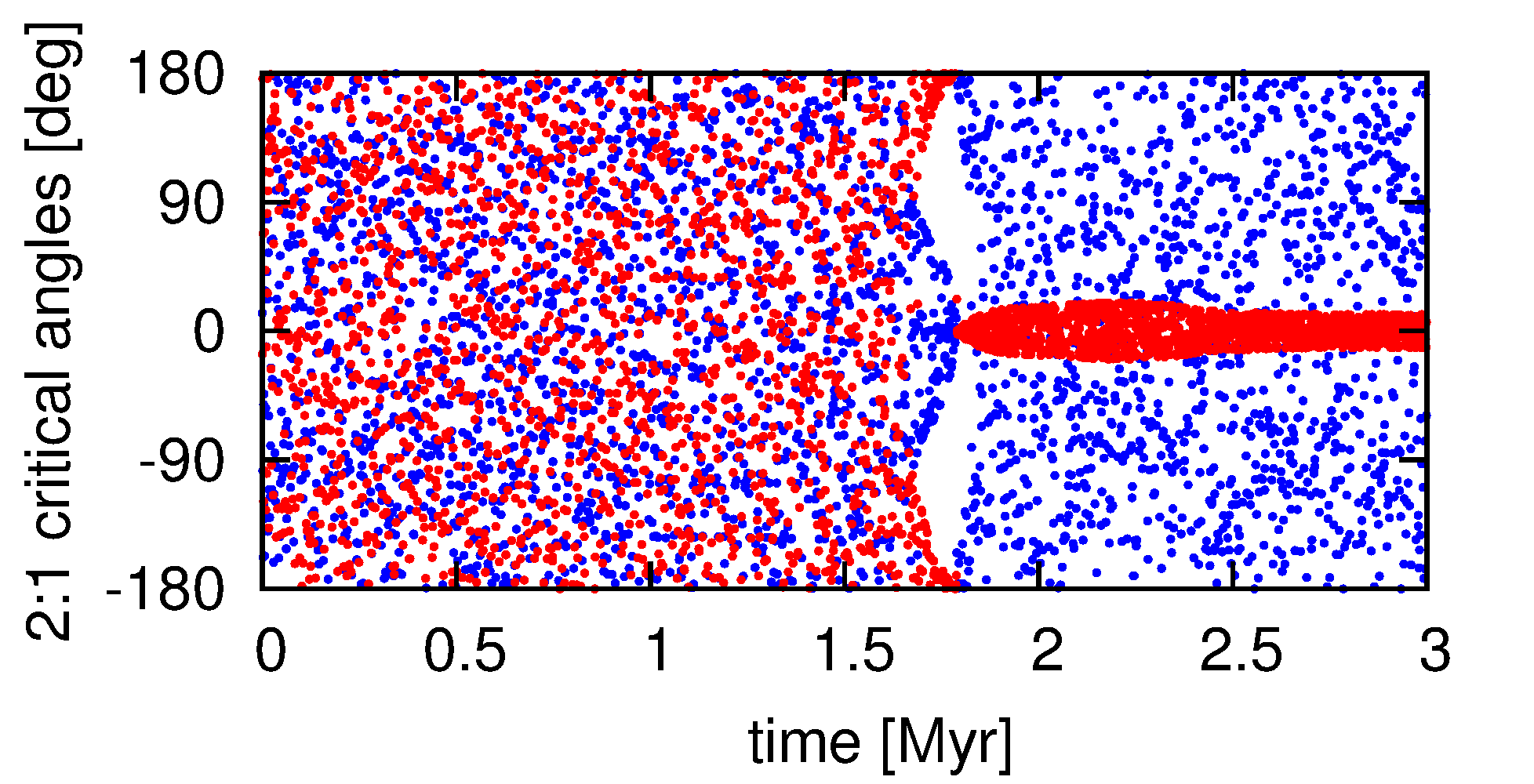}
}
}
}
\caption{Migration of example initial configurations which ended up in or close to 2:1~MMR illustrated in semi-major axes-, period ratio- and resonant angles-plots (from left-hand to right-hand column, respectively). Each row is for one system (they are enumerated with I, II, III and IV). The masses of the planets for each configuration are given in the left-hand column panels for the reference. The critical angles of the resonance are $\phi_{2:1}^{(1)}$ (red dots) and $\phi_{2:1}^{(2)}$ (blue).}
\label{fig:example_2_1}
\end{figure*}

\section{Periodic evolution}

We performed $3500$ simulations for different planets' masses and initial orbits. The masses of both planets are chosen from the range of $[2, 10]\,\mE$, initial $\log_{10} a_1[\au] \in [-0.1, 0.6]$ and initial $P_2/P_1 \in [1.3, 2.7]$. Distributions of all these quantities are uniform. For almost all the final systems, at least one of the resonant angles of a given MMR oscillates/librates, and if we take into account only the systems with $P_2/P_1 \lesssim 2.12$ (about half of the sample), for almost all the solutions both angles librate. Only the first order resonances are present. Even close to the nominal positions of higher order MMRs (5:3, 7:5, 8:5, etc.), the critical angles of these MMRs rotate. The evolution of a few representative systems which ended up in or close to 2:1~MMR are illustrated in Fig.~\ref{fig:example_2_1}.
Each row of this figure is for one solution, they are enumerated with I, II, III and IV, respectively. The left-hand, middle and right-hand columns show the evolution of the semi-major axes, the period ratio and the resonant angles, respectively. 

The first example (I) is a system which migrates convergently down to 2:1~MMR and stays there during the whole evolution. The period ratio is slightly higher than $2$, and the amplitudes of the resonant angles librations are very small. The first resonant angle of 2:1~MMR librates around $0$, while the second one librates around $180~$degrees.
System~II initially migrates convergently towards 3:2~MMR, stays close to the resonance for some time and then migrates out of this MMR. The final period ratio is $\sim 1.75$. At the end of the migration, both resonant angles of 2:1~MMR librate. The first one around $180~$degrees, while the second one around $0$. System~III initially migrates towards 3:2~MMR, next it moves out of the resonance. It passes through 2:1~MMR, reaches $P_2/P_1 \sim 3$ and returns. The final period ratio $\sim 2.1$. Both resonant angles librate, the first one around $0$, while the second one around $180~$degrees (like in system~I). It is a generic feature of all the systems originating from the migration within the model described in this paper. If a given system ends up with $P_2/P_1$ below the nominal value of a given resonance, the first critical angle (which is always the one with $\varpi_1$ in its definition) librates around $180~$degrees and the second one around $0$. If $P_2/P_1$ is larger than the nominal value, the situation is reversed.
System~IV ends up with $P_2/P_1 \sim 2.2$ and only the first angle oscillates, the second one rotates.

\begin{figure}
\centerline{
\vbox{
\hbox{
\includegraphics[width=0.24\textwidth]{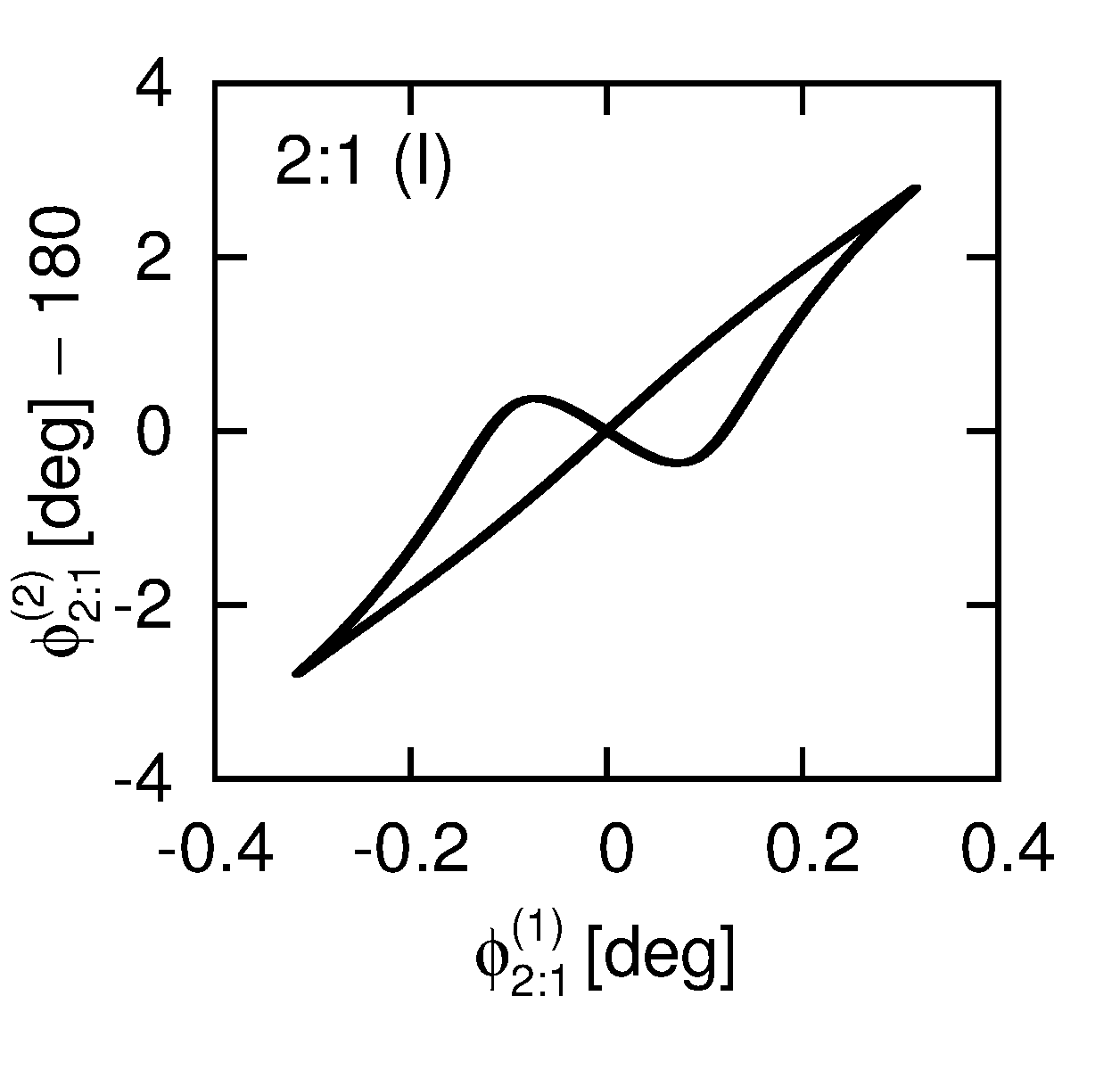}
\includegraphics[width=0.24\textwidth]{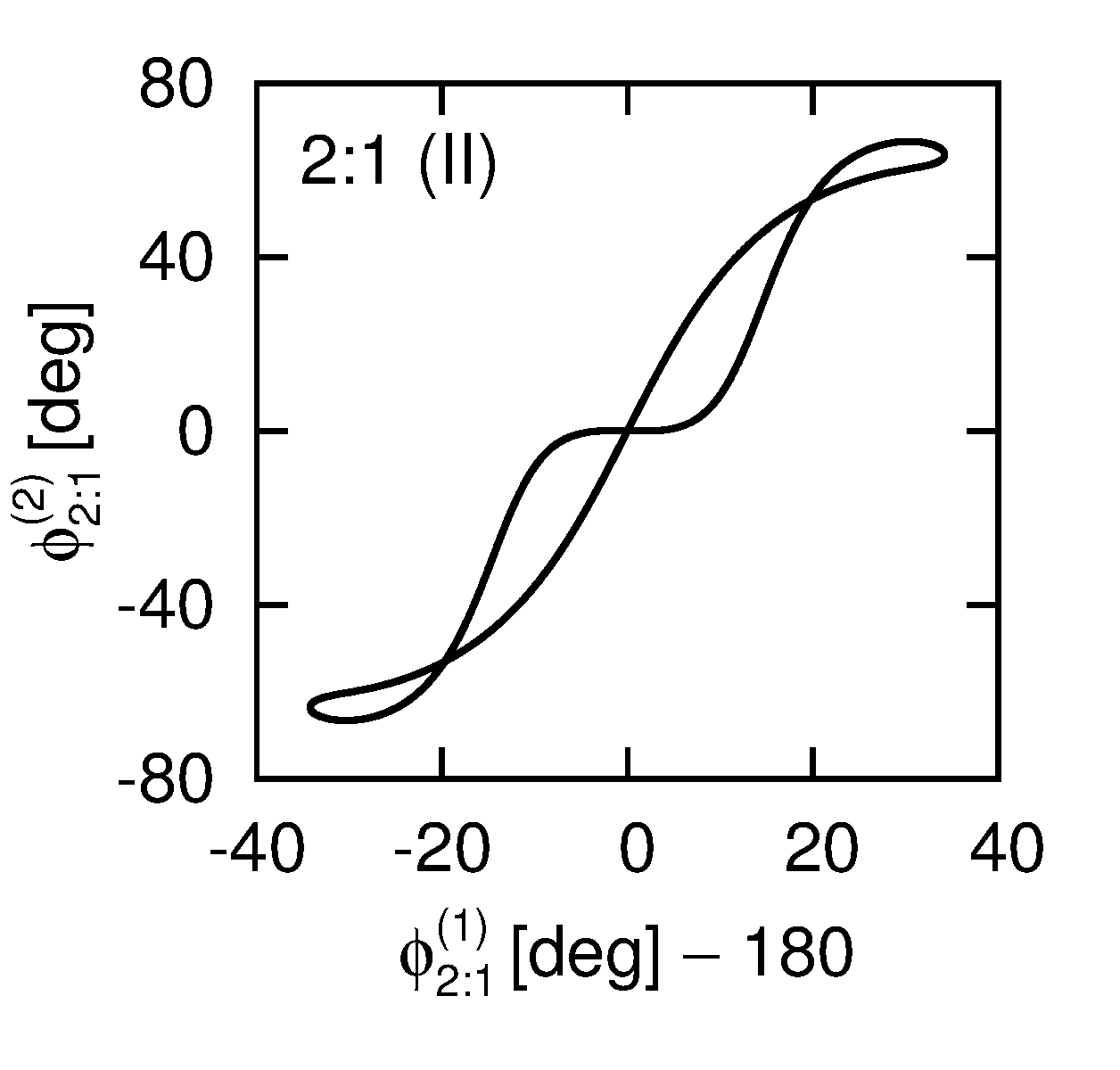}
}
\hbox{
\includegraphics[width=0.24\textwidth]{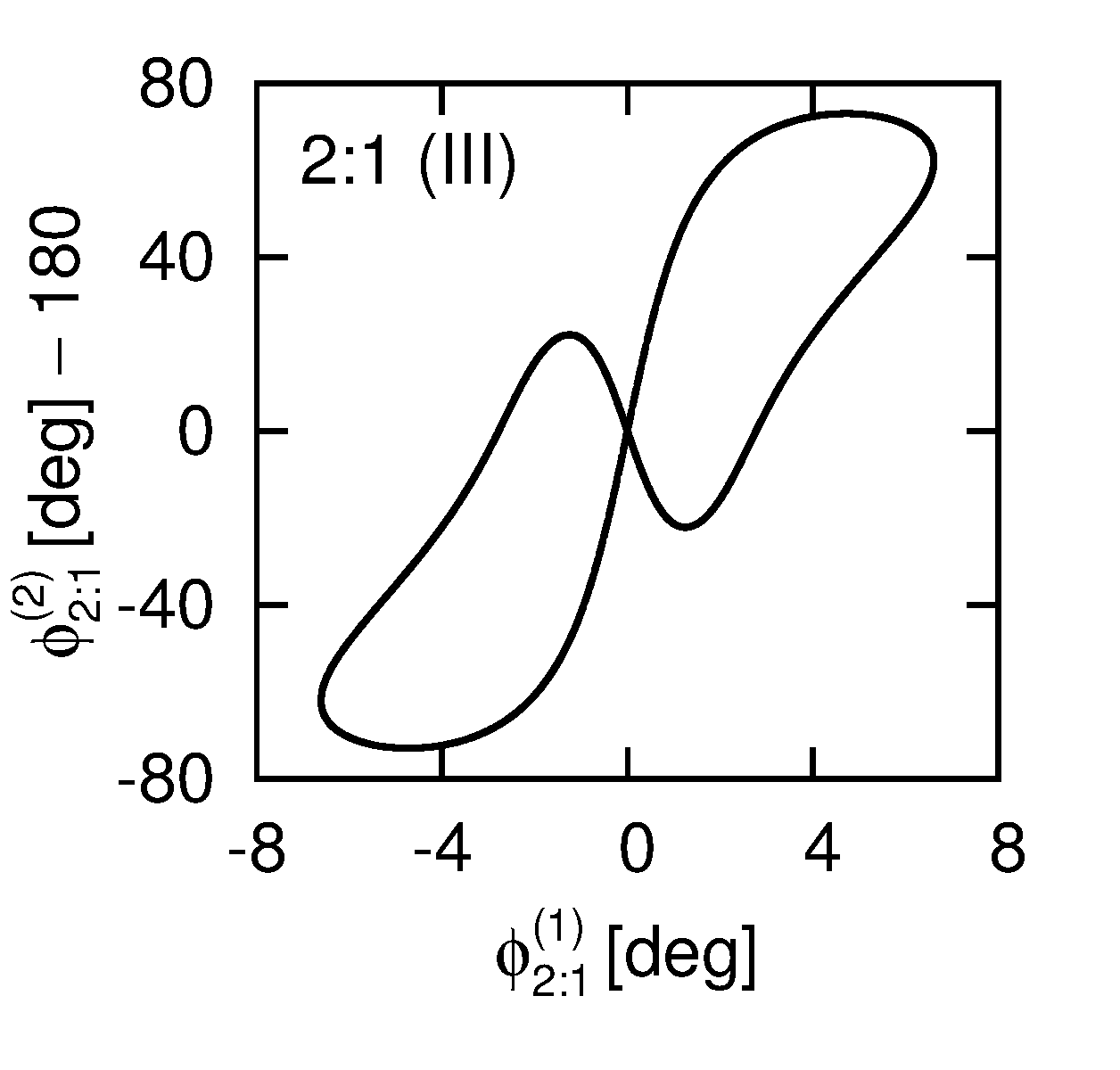}
\includegraphics[width=0.24\textwidth]{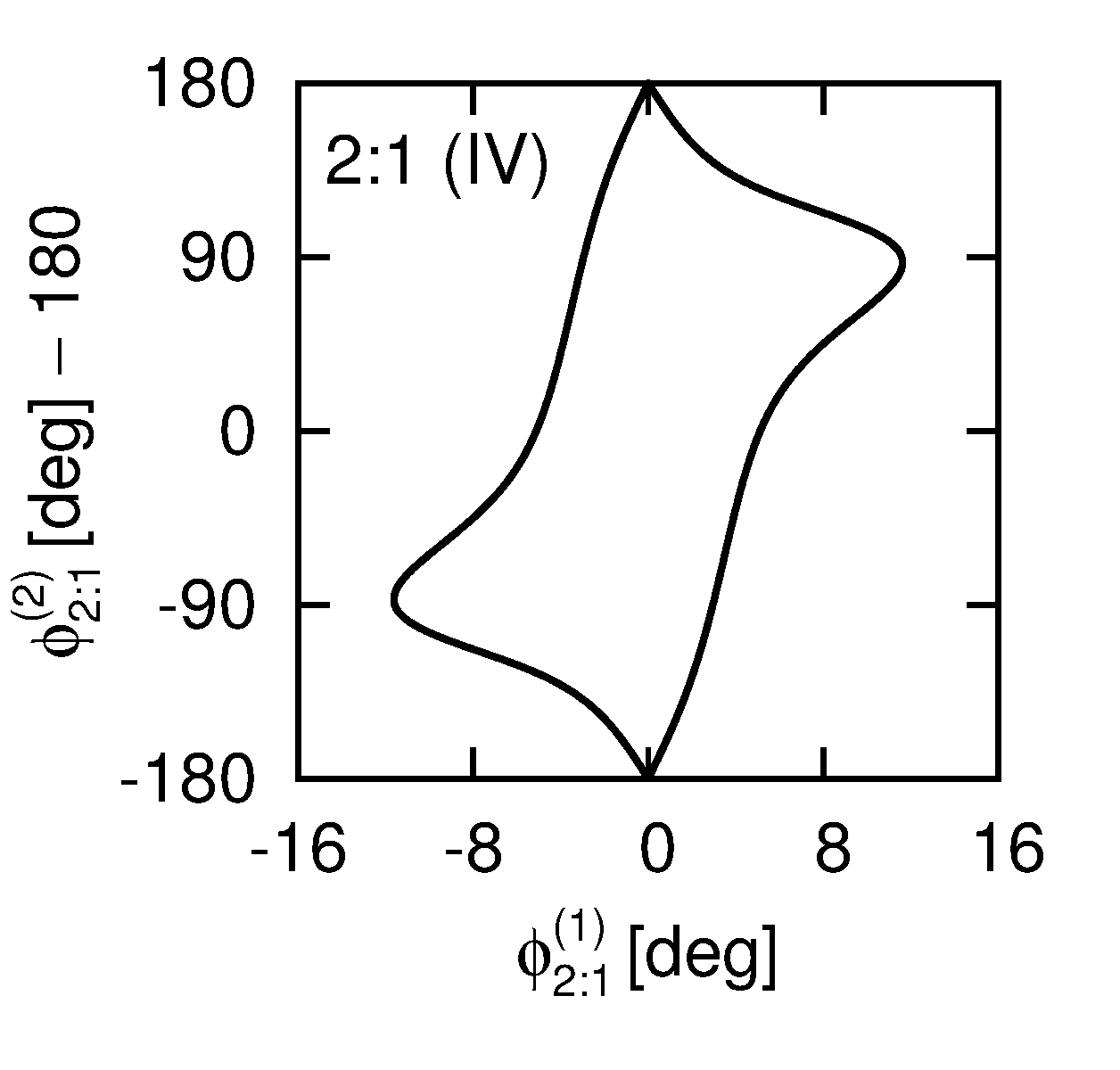}
}
}
}
\caption{Evolution of the resonant angles of 2:1~MMR. Each initial condition were integrated for $10^5 \times P_2$ within the $N$-body model of motion. Labels I, II, III and~IV corresponds to respective systems shown in Fig.~\ref{fig:example_2_1}.}
\label{fig:periodic_2_1}
\end{figure}

Almost all the solutions are characterized by oscilations/librations of at least one resonant angle. Another common feature of almost all the systems is that their evolution (after the disc is dispersed) is periodic. The final configurations whose migration were illustrated in Fig.~\ref{fig:example_2_1} were taken as initial conditions for the $N$-body simulations (see Fig.~\ref{fig:periodic_2_1}). The integration time is $10^5 \times P_2$. In all the cases, the evolution repeats every certain period. The period $\approx 30~$days for system~I (it is almost exactly the period of the outer planet, as both the apsidal lines move very slowly, i.e., with a period of $\sim 20\,\yr$).
The periods of the evolution of systems~II, III and~IV is shorter or longer than $P_2$, depending on if $P_2/P_1$ is smaller or larger than $2$. In all these cases the rotation of the apsidal lines is only a few times slower than the orbital motion of the outer planet. Eccentricities are also small $\sim 10^{-4}$ in these systems.
Similar behaviour of the angles can be observed for systems close to 3:2, 4:3 and 5:4~MMR.

\begin{figure}
\centerline{
\vbox{
\hbox{\includegraphics[width=0.49\textwidth]{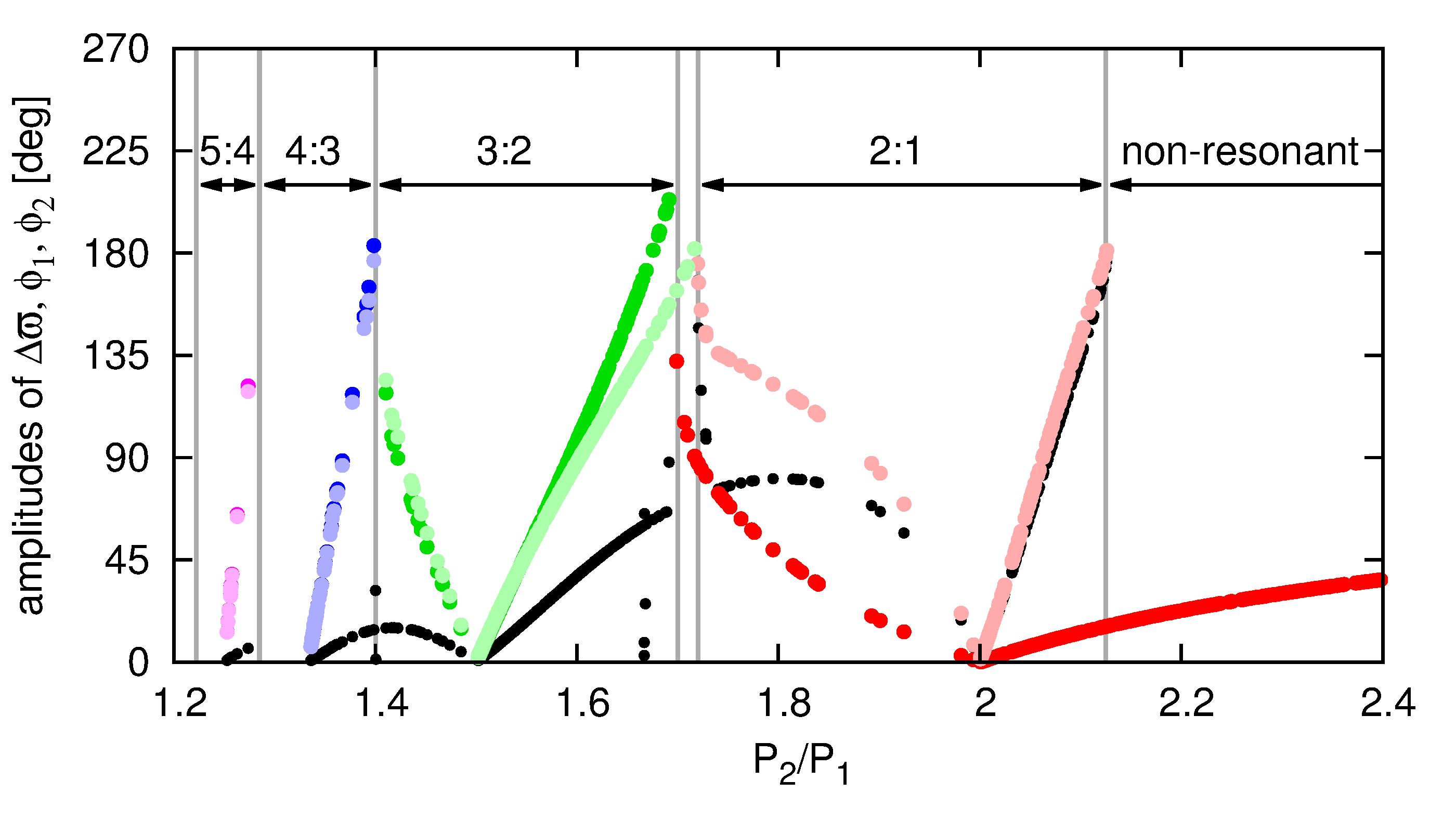}}
}
}
\caption{Amplitudes of $\Delta\varpi$ (black symbols) and the resonant angles of 2:1~MMR ($\phi_{2:1}^{(1)}$ -- red, $\phi_{2:1}^{(2)}$ -- light red), 3:2~MMR ($\phi_{3:2}^{(1)}$ -- green, $\phi_{3:2}^{(2)}$ -- light green), 4:3~MMR ($\phi_{4:3}^{(1)}$ -- blue, $\phi_{4:3}^{(2)}$ -- light blue) and 5:4~MMR ($\phi_{5:4}^{(1)}$ -- violet, $\phi_{5:4}^{(2)}$ -- light violet). Horizontal arrows show the ranges of first order MMRs.}
\label{fig:amplitudes}
\end{figure}

As we observed, the resonant angles of a given first order resonance (see Fig.~\ref{fig:example_2_1}) oscillate/librate around $0$ or $180~$degrees even for systems with $P_2/P_1$ significantly far from the nominal value of this MMR. It is possible because the apsidal motion for low eccentric systems may be very fast \citep{Batygin2013}. The eccentricities are getting lower and the amplitudes are getting higher when $P_2/P_1$ is moving away from a nominal value of a given MMR. 
Each system in the sample where integrated (within the $N$-body model) for $10^5 \times P_2$ taking as initial conditions the final orbital elements from the migration simulations. Figure~\ref{fig:amplitudes} shows how the amplitudes of librations of the resonant angles depend on the period ratio. The amplitudes for different angles are plotted with different colours (see the caption to Fig.~\ref{fig:amplitudes}). We can see that there are values of $P_2/P_1$ for which the amplitudes of two neighbouring MMRs tends to $\sim 180$~degrees. These values mean the borders between two neighbouring first order MMRs (they are marked with vertical lines). For $P_2/P_1 \in [\sim 1.7, \sim 1.72]$ there are solutions for which $\phi_{2:1}^{(1)}$ and $\phi_{3:2}^{(2)}$ oscillate. This range of the period ratio can be considered as the border between 2:1 and 3:2~MMRs.
Similar border can be seen between 3:2 and 4:3~MMRs (at $P_2/P_1 \sim 1.4$). A border between 4:3 and 5:4~MMRs cannot be noticed because there is no solutions in this range of $P_2/P_1$.
The outer border of 2:1~MMR takes place at $P_2/P_1 \sim 2.12$. The amplitude of the libration of $\phi_{2:1}^{(2)}$ tends to $\sim 180~$degrees for this value of the period ratio. In fact, for $P_2/P_1 \gtrsim 2.12$ the amplitude for $\phi_{2:1}^{(2)}$ equals $360~$degrees (it is not shown on the plot, as also the amplitudes for 3:2, 4:3 and 5:4~MMRs equals $360~$degrees in this range of the period ratio).
An interesting observation one can make by looking at the amplitudes diagram on Fig.~\ref{fig:amplitudes} is that all the systems are located along well defined curves. It means that the amplitudes of the librations of the resonant angles do not depend on the planets' masses.

\section{Limitations of the model}

There are several important assumptions of the model, which we list below. We plan to improve the model in future works.

\subsection{Disc model}

A natural assumption of 1+1D model is that the radial and the azimuthal structures of the disc can be studied separately. Moreover, we assumed that the turbulent viscosity is given by $\alpha$ prescription and that $\alpha$ is constant both in $r$ and in $z$-direction. If the turbulence results from the magnetorotational instability \citep[MRI][]{Balbus1991,Hawley1991}, the material of the disc needs to be sufficiently ionized to make this mechanism work efficiently. This condition is difficult to be fulfilled in cold dusty discs \citep{Blaes1994}. In such regions there might exist the so called ''dead zones'' \citep{Gammie1996a}, which do not transport the angular momentum. 
The ionization can occur due to collisions between particles, which is efficient for $T \gtrsim 1000\,\K$ and takes place near the star, $r \lesssim 0.1\,\au$ for relatively fast accretion rates of $10^{-9}$ to $10^{-8}\,\msun\,\yr^{-1}$ or even closer when the accretion is slower. Further from the star the galactic cosmic-ray ionization \citep{Gammie1996a} can be effective in regions of low optical depth (low density regions or close to the disc surface). In the latter case the midplane regions are neutral or only slightly ionized. In more realistic model $\alpha$ should be a function of $r$ and $z$ \citep[e.g.,][]{Kretke2010}.

\subsection{Planet-disc interaction}

There are three important assumption we made about the interaction between the disc and the planets. We assumed that the disc evolves as it was not affected by the planets. For planets of a few Earth masses and relatively massive disc ($\dot{M} \gtrsim 10^{-9}\,\msun\,\yr^{-1}$) the assumption is good. The formulae used in this paper takes into account only the fact that a planet locally disturb the disc, i.e., it opens a partial gap \citep[transition from Type~I to Type~II migration][]{Dittkrist2014}.
Second aspect is a single-planet approximation. One planet excites the waves in the disc and only this planets ''feels'' the gravitational force, which is a result of this perturbation. As shown in \citep{Podlewska-Gaca2012,Baruteau2013} this effect may be important, at least for planets which are able to open a partial gap in the disc. Another simplification is that all formulae used here were obtained for models with given $\gamma_1$ and $\gamma_2$ (the power indices of $\Sigma-$ and $T$-profiles), which were constant in sufficiently wide range of the disc (wide enough to encompass the wakes produced by the planet). Due to numerous transitions between the opacity regimes, a range of $r$ for which one can assume $\gamma_1$ and $\gamma_2$ to be constant may be narrower than the wake structure (in particular when the Lindblad torque is considered, as the corotation torque origins from a very tiny region around the position of the planet). On the other hand, the Lindblad torque is less sensitive to $\gamma_1$ and $\gamma_2$ than the corotation torque is.

\subsection{Planet mass growth}

Another important issue is the planet mass growth. As discussed in \citep{Bitsch2014} when a planet reach a trap (which works only for planets in certain mass range) it can stay there for some time, accrete material from the disc and finally pass the trap. Moreover, even if there is no trap in the disc, the evolution of a planet whose mass is not constant is, in general, different from the evolution of constant mass planet \citep[e.g.,][]{Alibert2013,Coleman2014,Machida2010}.
One could say that after some time the mass of the planet is almost constant. Still, one needs to start the integration from some initial orbits. On contrary, the model of growing mass is less arbitrary in this aspect, because one can start from earlier stage of the evolution, in which the choice of uniform distribution in $P_2/P_1$ and $\log_{10}a_1$ is better.

\section{Conclusions}

We have studied the migration of two super-Earths embedded in a protoplanetary disc. The forces acting on each planet are given by analytical prescriptions \citep{Paardekooper2011,Tanaka2004,Fendyke2014,Dittkrist2014}. It enabled us to study the orbital evolution of many initial configurations over the whole disc life-time. We showed that there are regions in the disc in which the migration is divergent. 
One of the regions corresponds to the outward migration zone, which is placed in the shadowed part of the disc (radii between $\sim 1$ and a few astronomical units) as well as in a transition zone between the irradiated and the shadowed parts of the disc. The second region of the divergent migration (at around of $0.5\,\au$) is a result of using more realistic opacity law \citep{Semenov2003} instead of a simpler law \citep{Ruden1991}. The third zone of the divergent migration ($r \lesssim 0.1\,\au$) corresponds to a region in which the opacity is dominated by gas. Other regions in the disc corresponds to the convergent migration.
In general, during the evolution, a given system spends some part of the time in regions of the convergent migration (the period ratio decreases and can be halted at one of the resonant values, i.e., $2/1$, $3/2$, $4/3$, etc.) and some other part of the time the system spends in regions of the divergent migration (the period ratio increases). The final configurations depends on the initial orbits as well as on the masses of the planets.

We performed $3500$ simulations for planets' masses and initial orbits chosen randomly. Almost all the systems ended up as periodic configurations. We found that the period ratio axis can be divided into several parts. The first part ($P_2/P_1 \gtrsim 2.12$) corresponds to non-resonant configurations, although one of two resonant angles of 2:1~MMR oscillates. Systems located in the second part of the period ratio axis ($P_2/P_1 \in [\sim 1.72, \sim 2.12]$) are involved in 2:1~MMR. Both resonant angles librate and the amplitudes of the librations depend on $P_2/P_1$. For $P_2/P_1 \approx 2$ the amplitudes are $\approx 0$. The amplitudes of both angles are monotonically increasing functions of $P_2/P_1$ when $P_2/P_1>2$ and monotonically decreasing functions of the period ratio when $P_2/P_1 < 2$. For $P_2/P_1 \sim 1.72$ the amplitudes of both angles reach $360~$degrees, which means that the system is located at the border of 2:1~MMR. For $P_2/P_1 \approx 2.12$, the amplitude of the second angle reaches $360~$degrees, while the amplitude of the first angle is still lower that $10~$degrees. The third part of the period ratios axis corresponds to 3:2~MMR. The range of period ratios of this resonance is $[\sim 1.4, \sim 1.7]$. Similarly to 2:1~MMR the amplitudes of the resonant angles librations are $\approx 0$ for $P_2/P_1 \approx 1.5$ and they increase when $P_2/P_1$ deviates from $1.5$ in either of the sides. At the borders of this resonance, the amplitudes reach $360~$degrees. Further parts of the period ratio axis corresponds to 4:3 and 5:4~MMRs. The amplitudes of the resonant angles librations have similar functional form as for the two previously discussed resonances. 

The general conclusion is that almost all the systems with $P_2/P_1 \lesssim 2.12$ are involved in one of the first order MMRs. Nevertheless, for systems whose period ratios are far from $(p+1)/p$, the ranges of $e \cos\varpi$ and $e \sin\varpi$, in which the resonant angles librate, are very small ($\sim 10^{-4} \div \sim 10^{-3}$). It means that even small perturbations of the orbits (e.g., due to turbulences in the disc or interactions with planetesimals) would probably change the behaviour of the resonant angles from librations to rotations and also move the system out of the periodic configuration.
The fact that most of the systems which underwent the smooth migration (convergent or divergent) evolve periodically might be tested with observations. If a given observed system was found to be close to a periodic orbit, it would mean that the system unlikely experienced strong perturbations at early stages of the evolution. This kind of studies could give the insight into the magnitude of turbulences in the disc as well as number and sizes of planetesimals.

\section*{acknowledgements}

Many thanks to Ewa Szuszkiewicz, Krzysztof Go\'zdziewski, Federico Panichi, Adam {\L}acny and Zijia Cui for discussions and comments on the manuscript. 
This work was supported by Polish National Science Centre MAESTRO grant DEC-2012/06/A/ST9/00276. The computations were performed on HPC cluster HAL9000 of the Computing Centre of the Faculty of Mathematics and Physics at the University of Szczecin, Reef cluster located
at the Pozna\'n Supercomputer and Network Centre PCSS (computational grant No. 195) and Zeus cluster which is a part of PLGrid Infrastructure (PRACE Distributed European Computing Initiative, DECI-11).

\bibliographystyle{mn2e}
\bibliography{ms}
\label{lastpage}
\end{document}